\documentclass[aps,prd,preprint,superscriptaddress,nofootinbib]{revtex4-1}
\usepackage{chay,graphicx}

\newcommand{\eps}{\epsilon}

\newcommand{\ip}{\ln \frac{\mu^2}{-p^2}}
\newcommand{\kp}{\mathbf{k}_{\perp}^2}
\newcommand{\kperp}{\mathbf{k}_{\perp}}
\newcommand{\bta}{\mbox{\boldmath $\eta_{\perp}$\unboldmath}\hspace{-0.1cm}}
\newcommand{\pperp}{\mbox{\boldmath $\mathcal{P}_{\perp}$\unboldmath}\hspace{-0.1cm}}

\begin{document}

\title{Structure of divergences in Drell-Yan process \\
with small transverse momentum}

\def\KU{Department of Physics, Korea University, Seoul 136-713, Korea} 

\def\Seoultech{Institute of Convergence Fundamental Studies \& School of Liberal Arts, Seoul National University of Science and 
Technology, Seoul 139-743, Korea}
\author{Junegone Chay}
\email[E-mail:]{chay@korea.ac.kr}
\affiliation{\KU}†
\author{Chul Kim}
\email[E-mail:]{chul@seoultech.ac.kr}
\affiliation{\Seoultech\vspace{0.5cm}} 

\begin{abstract} \vspace{0.1cm}\baselineskip 3.0 ex 
We consider the structure of divergences in Drell-Yan process with small transverse momentum.  
The factorization proof is not trivial because various kinds of divergences are intertwined in the collinear and soft parts 
at high orders.  We  prescribe a method to disentangle the divergences in the
framework of the soft-collinear effective theory.
The rapidity divergence is handled by introducing  the $\delta$ regulator in the collinear Wilson lines. 
The collinear part, which consists of the transverse-momentum-dependent parton distribution function 
(TMDPDF), is free of the rapidity divergence after the soft zero-bin subtraction.  There still remains the problem of  mixing
 between the ultraviolet and infrared divergences, which forbids the renormalization group description. We show that the mixing is 
cancelled by the soft function. This suggests that the collinear and soft parts should be treated as a whole in constructing a
consistent factorization theorem.  The renormalization group behavior of the combined collinear and soft parts is 
presented explicitly at one loop. 
We also show that the integrated PDF can be obtained by integrating the TMDPDF over
the transverse momentum.

\end{abstract}

\maketitle

\baselineskip 3.0 ex 

\section{Introduction}
Theoretical predictions of high-energy scattering rely on the factorization of scattering cross sections, in which
the hard, collinear and soft parts are separated to all orders in perturbation theory. Many inclusive cross 
sections have been proved to be factorized, and are used to compare with experiments. 
Less inclusive scattering processes are also of interest. Recently the transverse momentum 
distribution in the Higgs, $Z$ boson, and $t\overline{t}$ pair 
production  has redrawn new interest both theoretically 
\cite{Bozzi:2007pn,Bozzi:2008bb,Mantry:2009qz,Becher:2010tm,Becher:2011xn} and 
experimentally \cite{Abazov:2010kn,Aaltonen:2012fi}. 
But in high-energy scattering processes with nonzero momentum transverse to the 
beam direction,  the factorization proof becomes more involved. 
For example, in the Higgs production with the transverse momentum $q_T$ with 
$\Lambda_{\mathrm{QCD}} \ll q_T \ll M$, where $M$ is the Higgs mass, the scattering cross section involves unintegrated, or 
transverse-momentum-dependent parton distribution functions (TMDPDFs). The study of the transverse momentum 
distribution was first heralded by Ref.~\cite{Collins:1984kg}.
However, the factorization proof, which appears ubiquitously in high-energy processes including Drell-Yan process, has been 
recently under debate again.

Before we discuss many issues in proving factorization of the processes with transverse momentum, 
it is worth delineating why the Drell-Yan process with small transverse momentum is so intricate. When we consider radiative 
corrections in massless gauge theories, there appear ultraviolet (UV) and infrared (IR) divergences.  The
UV and IR divergences should appear separately with no mixing in order for the theory to be consistent. 
In many inclusive processes, this was shown explicitly in the collinear and soft parts. In the process where  the transverse
momentum for a final-state particle is fixed, the UV divergence to be obtained by integrating over the transverse momentum 
does not emerge yet.
Therefore the separation of the UV and IR divergences seems murky at first sight. Furthermore, when the transverse momentum is fixed,
there appears another type of divergence called rapidity divergence, or lightcone singularity. The issues in the Drell-Yan processes with
small transverse momentum reside in the appropriate treatment of rapidity divergence and in finding an adequate formalism in which
the separation of the UV and IR divergences is established.  

The  first issue is the existence of rapidity divergence, which has been known for a long time \cite{Collins:2003fm},  
and various ways of handling it have been suggested. 
The source of the rapidity divergence can be understood by the following argument:
In QCD, the Kinoshita-Lee-Nauenberg theorem \cite{Kinoshita:1962ur,Lee:1964is} guarantees that the divergence 
in virtual corrections at a given order in the strong coupling $\alpha_s$ is cancelled by the divergence in real gluon emissions, 
thus rendering inclusive scattering cross sections free of infrared divergence. The presence of infrared divergences can be traced by 
considering the kinematics in each case. In virtual corrections, the loop momentum can be 
collinear to an energetic particle, or soft, which results in collinear or soft divergences, or it can be both collinear and soft.
In real emissions, the emitted gluon can also be collinear to an energetic particle or soft, or both as long as the phase space allows it.
 These divergences cancel in inclusive processes, and the inclusive scattering cross sections are physically meaningful in the absence
of the collinear and soft divergences. However, if we consider differential cross sections with 
fixed transverse momentum, this cancellation becomes incomplete. That is, when the transverse momentum of an emitted 
gluon is fixed,  the phase space of the emitted gluon does not cover all the available phase space in virtual corrections, and the IR
divergences do not cancel because of the imbalance of the phase space between real and virtual corrections. 
The presence of the infrared divergence due to this 
incomplete cancellation is referred to as the ``rapidity divergence'' or lightcone singularity since it is the result caused 
by the interaction of the energetic particle 
with collinear gluons with infinite rapidity. The dimensional regularization cannot regulate the rapidity divergence.

The difficulty in regulating the rapidity divergence was first noticed by Collins \cite{Collins:2003fm}. 
Collins and  Soper \cite{Collins:1981uk,Collins:1981va} suggested an intriguing idea of defining the operator whose matrix 
elements generate TMDPDF by tilting the Wilson lines which connect the fermion fields to make the operator gauge invariant. 
In association with this tilting, they introduced 
an additional scale of which the TMDPDF should be independent, and the corresponding evolution equation was developed.   
This method, while effective in handling the rapidity divergence, has a nuisance. First, the dependence of the additional scale 
appears in the hard, collinear and soft parts, which troubles the factorization proof. Though the dependence of the additional scale 
disappears when all the parts are summed, it is not clear how to approach the physical limit, that is, the limit in which all 
the particles are put on the lightcone. Finally it is not straightforward to obtain the integrated PDF from the TMDPDF by 
integrating over the transverse momentum  \cite{Ji:2004wu}.

Chiu et al. \cite{Chiu:2011qc,Chiu:2012ir} developed the idea of rapidity renormalization to handle the rapidity divergence. 
In high-energy scattering, 
there are collinear particles and soft particles, and they note that particles with the same offshellness can be labeled as collinear or soft 
depending on their rapidities. Therefore 
a rapidity scale is set up to distinguish collinear and soft particles. However, since the physics is 
independent of this arbitrary scale for the 
separation, the evolution equations of the collinear and the soft parts can be obtained with respect to the scaling of this rapidity scale. 
Technically they modified collinear and soft Wilson lines in such a way that Wilson lines depend on the rapidity scale. The rapidity scale 
dependence of the Wilson line
is determined to satisfy the requirement that the cross section be independent of the rapidity scale, that is, the rapidity scale dependence 
of the collinear part is cancelled by that of the soft part. 
Though the motivations come from different reasoning, the approaches of Collins et al. \cite{Collins:1984kg} and 
Chiu et al. \cite{Chiu:2011qc,Chiu:2012ir} share in common 
the fact that they introduced an additional scale of which physical results should be independent. This is reflected in the evolution 
equations of the collinear and soft parts.  The purpose of this approach is also focused to handle the rapidity divergence.

In this paper, a straightforward way of regulating the rapidity divergence
is presented in the framework of the soft-collinear effective theory (SCET) \cite{Bauer:2000ew,Bauer:2000yr,Bauer:2001yt}.
The regularization for the rapidity
divergence is achieved by introducing the $\delta$ regulator in the collinear Wilson lines  only and not in any other propagators.
Chiu et al. \cite{Chiu:2009yx}  introduced the  $\delta$ regulator, which is an infrared regulator inserted in every propagator,
as well as in the Wilson lines. They showed that the $\delta$ dependence of the collinear part 
in the back-to-back current is cancelled by the zero-bin subtraction. They stressed the importance 
of the proper zero-bin subtraction in the cancellation of the $\delta$ dependence. The $\delta$ regulators can be  given
as arbitrary parameters, or they can be related to the offshellness of the external particles. In either case,
the $\delta$ dependence in the collinear and soft parts is different and cannot be cancelled unless a specific relation is
imposed.  A simple relation can be found in a special case like a back-to-back current  when the offshellness is employed
to regulate divergences.
However, when multijets are involved, $\delta$ depends on the offshellness of all the external particles in a complicated way.
The collinear Wilson line in the $n$ direction is obtained by considering the emission of $n$-collinear gluons from the other particles 
not in the $n$ direction. Then the intermediate states become offshell, and the collinear Wilson line is obtained by integrating out
this offshellness, and taking the leading term. If a back-to-back current is involved only, the $\delta$ regulator can be related to 
the offshellness of the particle in the $\overline{n}$ direction for the $n$-collinear Wilson line. If we consider a process in which
there are many jets, the offshellness in a collinear Wilson line depends on all the other jets in the process. On the other hand, the $n$-soft
Wilson line is obtained by considering  soft gluon emissions from the $n$-collinear particle. Therefore if the offshellness 
is included, 
the soft Wilson line involves only the offshellness of the corresponding collinear particle. 
Here the rapidity divergence is handled by inserting the $\delta$ regulator in the collinear Wilson lines only. Other regulators,
though similar to the $\delta$ regulator in the collinear Wilson lines in form, regulate IR divergences.

The second issue is to separate the remaining UV and IR divergences. The presence of the mixing between the UV and IR
divergences is troublesome because the physics at a high scale and a low scale is mixed, which does not make sense. 
Therefore the decoupling of the UV and IR divergences in loop calculations is essential to guarantee the consistency of the theory. 
In contrast to inclusive processes, the appropriate combination in which this decoupling occurs is still vague. It makes  
the definition of the TMDPDF itself  controversial. The problem is whether the soft contribution or part of it should 
be included in the matrix element of the collinear operators which comprise the backbone of the TMDPDF. 
The underlying idea is to find a suitable combination of collinear and soft parts
such that UV and IR divergences at higher orders are separated in order to prove factorization. 
A specific combination of collinear and soft parts has been claimed to achieve this separation  \cite{Collins:2011zzd}.  
But once soft part is included in the TMDPDF, there follows another issue, the universality of the TMDPDF. 
The soft part interacts with all the different collinear sectors, and it is different
for different final hadronic states. Therefore the definition of the TMDPDF in one process may lose its meaning 
in other processes once the soft part in a specific process is included in the TMDPDF though the TMDPDFs in 
two different processes can be related to each other. 
This kind of relation for the PDF in deep inelastic scattering and Drell-Yan processes has been discussed 
when the soft part is included  in the PDF \cite{Sterman:1986fn}.

In Ref.~\cite{GarciaEchevarria:2011rb},  an intriguing method to define the factorized TMDPDF was developed. 
First, the authors put all the particles on the lightcone, which
is physical. Therefore there is no need to introduce additional scales due to the tilting of Wilson lines or rapidity separation. 
Obviously a technique to handle the rapidity divergence is necessary, and the $\delta$ regulator, slightly different from
that in Ref.~\cite{Chiu:2009yx}, is employed. 
They defined the TMDPDF in terms of the collinear part with the soft function, which is similar to the approach 
of  Collins \cite{Collins:2011zzd}.
The reason  is that the $\delta$ dependence, which indicates the presence of the 
rapidity divergence, is cancelled only through a specific combination of the collinear and  soft parts as they claimed. It is based 
on the observation \cite{Chay:2005rz,Idilbi:2007ff,Idilbi:2007yi} that the zero-bin contribution can be identified as the soft part.
It is true for back-to-back currents to set up a relation of the regulators in the collinear and soft parts, but in general it does not hold. 
For example, if there are many jets or heavy colored particles in the process,  there is no simple relation between them.

In this paper, the TMDPDF is defined in terms of the collinear fields only, thus it is universal in all high-energy processes. 
All the particles are put on the lightcone and the rapidity divergence is handled by the $\delta$ regulator in the collinear Wilson lines. 
 The collinear part itself along with the 
corresponding zero-bin subtraction is independent of the $\delta$ regulator. The zero-bin contribution is performed in the limit where
collinear particles become soft (not ultrasoft) to avoid double counting \cite{Manohar:2006nz}.
The infrared divergence from the soft Wilson lines is cancelled
by that of the collinear part, which is controlled by the offshellness of the external particles.  Combining all the ingredients, we obtain
a factorized form of the scattering cross section with small transverse momentum in Drell-Yan processes. The hard part is the 
Wilson coefficients obtained in matching the current operators between QCD and SCET. The 
collinear part consists of the TMDPDF, and the remaining part is the soft function. We develop a  method to handle 
the rapidity divergence,
with the decoupling of the UV and IR divergences, and show the result explicitly at next-to-leading order (NLO) accuracy.

The structure of the paper is as follows: In Section \ref{facdy}, the factorized form of the scattering cross section in 
Drell-Yan process is presented.  In Section \ref{tmdpdf}, the 
TMDPDF is defined as the matrix elements of collinear fields, and the one-loop  corrections  are computed. 
The role of the zero-bin subtraction to avoid double counting 
is discussed and we argue that the soft zero-bin subtraction is the appropriate method for small transverse momentum 
$q_T \sim Q\lambda$.  
 In Section~\ref{softone}, the soft function is defined in terms of the soft Wilson lines, and its one-loop 
correction is presented. The relation between the TMDPDF and the integrated PDF is discussed in Section~\ref{comp}, and we 
explain how the integrated PDF is obtained from the TMDPDF. In Section~\ref{renor}, the renormalization group behavior
of the scattering cross section at NLO is presented. In Section~\ref{decuvir}, the decoupling of the UV and IR divergences is 
elucidated in detail. And finally in Section~\ref{conc}, we summarize the procedure for
taming the rapidity divergence  with the decoupling of the UV and IR divergences, and give a conclusion. 
In Appendix, the properties of the $\mu^2$ distribution and the
infinity distribution are described. 

\section{Factorization in Drell-Yan process\label{facdy}}
Let us consider  Drell-Yan process 
$p \overline{p} \rightarrow \ell^+ \ell^- +X$, where the lepton pair $\ell^+ \ell^-$ is produced with small transverse momentum $q_T$
($\Lambda_{\mathrm{QCD}} \ll q_T \ll Q$). The invariant mass of the lepton pair $Q$ is the large scale.  The
incoming particles are $n$-collinear and $\overline{n}$-collinear, and their momenta scale as
\begin{equation}
p_n^{\mu} =(\overline{n}\cdot p_n, n\cdot p_n, p_{n\perp}) \sim Q(1,\lambda^2, \lambda), \ 
p_{\bar{n}}^{\mu} =(\overline{n}\cdot p_{\bar{n}}, n\cdot p_{\bar{n}}, p_{\bar{n}\perp}) \sim Q(\lambda^2, 1,\lambda),
\label{dxsec}
\end{equation}
where $\lambda =q_T/Q$ is the small parameter. We also denote $\overline{n}\cdot p =p^-$, $n\cdot p =p^+$.
There are soft particles, whose momenta scale as
$p_s^{\mu} \sim Q(\lambda, \lambda,\lambda)$, in the final state to balance the transverse momentum of the 
lepton pair. In SCET, the interactions between collinear and soft particles are not allowed since they put each particle far off their mass shells. 
Therefore the soft part is decoupled and is expressed in terms of the soft Wilson lines.
The ultrasoft (usoft) particles can also exist, but they do not contribute to the cross section with transverse momentum 
$q_T\sim Q\lambda$.

The differential scattering cross section  can be written  as
\begin{equation}
d\sigma = \frac{\sigma_0 (q^2)}{s} \frac{d^4 q}{(2\pi)^4} \int d^4 x e^{-i q\cdot x} (-g^{\mu\nu} N_c) \langle N_1 N_2|
J_{\mu}^{\dagger} (x) J_{\nu} (0)|N_1 N_2\rangle,
\end{equation}
where $\sigma_0 = 4\pi \alpha^2 Q_f^2/(3q^2 N_c)$, with the electric charge $Q_f$ and the number of colors $N_c$. 
The electromagnetic current is given by 
$J_{\mu} = \overline{f} \gamma^{\mu} f$. The sum over the flavors $f$  is implied. 
We consider only the electroproduction here, but the
weak interaction can be easily implemented. 
We choose the direction of the momentum $P_1$ for the nucleon $N_1$ to be in the $n$ direction, and 
that of $P_2$ for $N_2$ in the $\overline{n}$ direction.
In SCET, the current operator for the quark-antiquark annihilation can be written as
\begin{equation} \label{current}
J_{\mu} (x) = C(Q) e^{-i(\bar{n}\cdot p n\cdot x/2 + n\cdot \bar{p} \bar{n}\cdot x/2) } 
\overline{\chi}_{n,p} Y_n^{\dagger} \gamma_{\mu}Y_{\bar{n}} 
\chi_{\bar{n}, \bar{p}} (\mathbf{x}_{\perp}),
\end{equation}
where $Q^2 = q^2$, and $\chi_n = W_n^{\dagger} \xi_n$ is the gauge-invariant collinear fermion field 
with the collinear Wilson line $W_n$.
 $C(Q)$ is the Wilson coefficient obtained in matching between the full
theory and SCET.  Unlike the conventional SCET formalism in which the momenta of order 
$Q$ and $Q\lambda$ are label momenta, 
only the momenta of order $Q$ are the only label momenta to be extracted in Eq.~(\ref{current}). It is because the soft momentum of 
order $Q\lambda$
is the dynamical degree of freedom, and the size of the fluctuation after extracting label momenta is of order $\mathbf{x}_{\perp} \sim 
(Q\lambda)^{-1}$. 

The differential cross section in SCET can be written as
\begin{eqnarray}
d\sigma &=& \frac{d^4 q}{(2\pi)^4} \frac{\sigma_0}{s} (-g^{\mu\nu}N_c) \int d\omega_1 d\omega_2 
\int d^4 x e^{-i q\cdot x} e^{i(\omega_1 n\cdot x/2 +\omega_2 \bar{n}\cdot x/2)} |C(Q)|^2 \nonumber \\
&\times& \langle N_1 (P_1) N_2 (P_2)| \Bigl[\overline{\chi}_n\delta (\omega_1 -\overline{n}\cdot \mathcal{P}^{\dagger}) 
\Bigr] Y_n^{\dagger} 
\gamma_{\perp \mu} Y_{\bar{n}} \Bigl[\delta(\omega_2 +n\cdot \mathcal{P})\chi_{\bar{n}}  (\mathbf{x}_{\perp}) \Bigr] 
\nonumber \\
&&\times \overline{\chi}_{\bar{n}} Y_{\bar{n}}^{\dagger} \gamma_{\perp \nu}
Y_n \chi_n (0) |N_1 (P_1)  N_2 (P_2)\rangle,
\end{eqnarray}
where $\omega_1 = x_1 \overline{n} \cdot P_1$,  $\omega_2 = x_2 n \cdot P_2$, and $x_1$, $x_2$ are the longitudinal momentum
fractions of the incoming partons. Here 
$\overline{n}\cdot \mathcal{P}$, $n\cdot\mathcal{P}$ are the label momentum operators for the particles in the 
$n$ and $\overline{n}$ directions respectively and the operators are applied inside the brackets.
Since the soft Wilson lines are decoupled from collinear particles, they can be extracted out, and expressed in terms of the vacuum 
expectation values.
The fields at $\mathbf{x}_{\perp}$ can be expressed by the fields at the origin as
\begin{eqnarray}
\Bigl(\overline{\chi}_n\Bigr)_{\alpha}^a (\mathbf{x}_{\perp}) &=&\int d^2 \mathbf{k}_{1\perp}  
e^{-i \mathbf{k}_{1\perp} \cdot \mathbf{x}_{\perp}}
\Bigl[ \Bigl(\overline{\chi}_n\Bigr)_{\alpha}^a  (0) \delta^{(2)} (\mathbf{k}_{1\perp} -\pperp^{\dagger}) \Bigr],  \nonumber \\
\Bigl(\chi_{\bar{n}}\Bigr)_{\alpha}^a (\mathbf{x}_{\perp}) &=& \int d^2 \mathbf{k}_{2\perp}  
e^{-i \mathbf{k}_{2\perp} \cdot \mathbf{x}_{\perp}}
\Bigl[  \delta^{(2)} (\mathbf{k}_{2\perp} +\pperp) \Bigl(\chi_{\bar{n}}\Bigr)_{\alpha}^a  (0)\Bigr], \nonumber \\
(Y_n^{\dagger} Y_{\overline{n}})_{ab} (\mathbf{x}_{\perp}) (Y_{\bar{n}} Y_n)_{cd} (0)
&=& \int d^2 \bta e^{i\bta \cdot \mathbf{x}_{\perp}} (Y_n^{\dagger} Y_{\bar{n}})_{ab} 
\delta^{(2)} (\bta +i\nabla_{\perp}) (Y_{\bar{n}}^{\dagger} Y_n)_{cd} (0),
\end{eqnarray}
where $\pperp$ is the operator extracting the transverse momentum.  The TMDPDFs are defined after taking the spin average as
\begin{eqnarray} \label{deftmdpdf}
\langle N_1 | (\chi_n)_{\alpha}^a \Bigl[(\overline{\chi}_n)_{\beta}^b \delta \Bigl( x_1 
-\frac{\overline{n} \cdot \mathcal{P}^{\dagger}}{\overline{n}\cdot P_1} \Bigr)  \delta^{(2)} (\mathbf{k}_{1\perp} - 
\pperp^{\dagger})\Bigr] 
 |N_1 \rangle &=& \frac{\overline{n}\cdot P_1 }{2N_c} 
\delta^{ab} \Bigl(\frac{\FMslash{n}}{2} \Bigr)_{\alpha\beta} f_{q/N_1} (x_1, \mathbf{k}_{1\perp}), \nonumber \\
\langle N_2|\Bigl[\delta \Bigl( x_2 +\frac{n\cdot \mathcal{P}}{n\cdot P_2}\Bigr) \delta^{(2)} (\mathbf{k}_{2\perp}
+ \pperp) (\chi_{\overline{n}})_{\alpha}^a \Bigr] (\overline{\chi}_{\overline{n}})_{\beta}^b |N_2\rangle &=&
\frac{n\cdot P_2}{2N_c} \delta^{ab} \Bigl(\frac{\FMslash{\overline{n}}}{2} \Bigr)_{\alpha\beta} 
f_{\overline{q}/N_2} (x_2, \mathbf{k}_{2\perp}).
\end{eqnarray}
After straightforward algebra, the differential scattering cross section can be written as
\begin{eqnarray} \label{xsec}
\frac{d\sigma}{d^2\mathbf{q}_{\perp}} &=& \int dx_1 dx_2 \sigma_0 H(Q^2,\mu) \int d^2 
\mathbf{k}_{1\perp} d^2 \mathbf{k}_{2\perp}
d^2 \bta \delta^{(2)} ( \mathbf{k}_{1\perp} +\mathbf{k}_{2\perp} -\bta -\mathbf{q}_{\perp}) \nonumber \\
&\times& f_{q/N_1} (x_1, \mathbf{k}_{1\perp}) f_{\overline{q}/N_2} (x_2,\mathbf{k}_{2\perp}) S(\bta),
\end{eqnarray}
where $s=\overline{n}\cdot P_1 n\cdot P_2$,  $Q^2 = x_1x_2 s$, and $\mathbf{q}_{\perp}$ is the transverse momentum
of the lepton pair.  From  Eq.~(\ref{deftmdpdf}), the TMDPDFs are given as
\begin{eqnarray} \label{def}
f_{q/N_1}  (x,\mathbf{k}_{\perp} )&=& \langle N_1 (P_1)| \overline{\chi}_n \frac{\FMslash{\overline{n}}}{2} \delta^{(2)}
 (\mathbf{k}_{\perp} - \pperp)\delta \Bigl( \overline{n}\cdot \mathcal{P} -x\overline{n}\cdot P_1\Bigr) 
\chi_n |N_1 (P_1)\rangle, 
\nonumber \\
f_{\overline{q}/N_2} (x,\mathbf{k}_{\perp} )&=& \langle N_2 (P_2)| \overline{\chi}_{\overline{n}} 
\frac{\FMslash{n}}{2} \delta^{(2)} (\mathbf{k}_{\perp} + \pperp)\delta 
\Bigl( n\cdot \mathcal{P} +x n\cdot P_2  \Bigr) \chi_{\overline{n}} |N_2 (P_2)\rangle, 
\end{eqnarray}
and the soft Wilson line is given by
\begin{equation} \label{swilson}
S(\bta) =\frac{1}{N_c}\langle 0| \mathrm{tr}\Bigl[  Y_n^{\dagger} Y_{\overline{n}} \delta^{(2)} (\bta + i\nabla_{\perp}) 
Y_{\overline{n}}^{\dagger} Y_n \Bigr]|0\rangle. 
\end{equation}
There is an alternative definition of TMDPDF in previous literature in terms of the delta functions of the transverse momentum 
in $D-2$ dimensions instead of two dimensions. With this definition, the radiative corrections should change accordingly, but
equivalent results can be obtained. However, the definition in Eq.~(\ref{def}) is preferred in this paper since the TMDPDF is an observable
and it should be defined in four spacetime dimensions. If the Wilson lines $Y_n$ and $Y_{\overline{n}}$ are usoft, the momentum
 fluctuation is of order $Q\lambda^2$ and the momentum operator  in the delta function in Eq.~(\ref{swilson}) can
be put to zero at leading order, and the usoft Wilson lines cancel. The fact that usoft interactions do not contribute to the
process with the transverse momentum of order $Q\lambda$ to all orders in $\alpha_s$ 
manifests itself due to the cancellation of the usoft Wilson lines.

Eq.~(\ref{xsec}) is the factorized form for the differential scattering cross section. The hard function 
$H(Q) =|C(Q)|^2$ is the Wilson coefficient of the 
current, the collinear part consists of the product of two TMDPDFs, and the soft Wilson lines comprise the soft part. 
The radiative corrections of the hard, collinear and soft parts can be computed
separately in perturbation theory. 
However, the infrared divergences, or the dependence of the regulators in actual calculations are intertwined between the collinear
and soft parts, of which the disentanglement is the main topic of this paper.  It will be shown that the dependence of the 
$\delta$ regulator, which controls the rapidity divergence, cancels in the collinear sector. The mixing between the UV and IR divergences
disappears when the virtual and real corrections are added in the collinear sector and in the soft sector. It also disappears when
the collinear and soft parts are added in the virtual corrections and in the real gluon emissions. This observation implies the separation
of divergences to all orders. 

The TMDPDF can be expressed in terms of the integrated PDF $\phi_{q/N} (z)$ as
\begin{equation} \label{tmdint}
f_{q/N} (x, \kp, \mu) = \int_x^1 \frac{dz}{z} J_{qq}(z, \kp,\mu) \phi_{q/N} \Bigl(\frac{x}{z},\mu\Bigr) =
\int_x^1 \frac{dz}{z} J_{qq}\Bigl(\frac{x}{z}, \kp,\mu \Bigr) \phi_{q/N} (z,\mu).
\end{equation}
This corresponds to the matching of the operator for the TMDPDF at the scale $\sim Q\lambda$ to the operator for the integrated PDF
at the scale $\sim Q\lambda^2$, and the kernel $J_{qq}(z,\kp,\mu)$ is the matching coefficient. 
The differential cross section in Eq.~(\ref{xsec}) can be expressed in terms of the integrated PDF as
\begin{eqnarray} \label{vir}
\frac{d\sigma}{d^2\mathbf{q}_{\perp}} &=& \int dx_1 dx_2\sigma_0 H(Q^2,\mu) \int d^2 
\mathbf{k}_{1\perp} d^2 \mathbf{k}_{2\perp}
d^2 \bta  \delta^{(2)} ( \mathbf{k}_{1\perp} +\mathbf{k}_{2\perp} -\bta -\mathbf{q}_{\perp}) S(\bta,\mu) \\
&\times&  \int_{x_1}^1 \frac{dz_1}{z_1} \int_{x_2}^1 \frac{dz_2}{z_2} J_{qq}(z_1, \mathbf{k}_{1\perp},\mu) 
J_{\bar{q}\bar{q}}(z_2, \mathbf{k}_{2\perp},\mu) \phi_{q/N_1} \Bigl(\frac{x_1}{z_1},\mu\Bigr) \phi_{\bar{q}/N_2} 
\Bigl(\frac{x_2}{z_2},\mu\Bigr)  \nonumber \\
&=&\int dx_1 dx_2  \sigma_0 H(Q^2,\mu)  \int_{x_1}^1\frac{dz_1}{z_1} 
\int_{x_2}^1 \frac{dz_2}{z_2} W(z_1,z_2, \mathbf{q}_{\perp},\mu) 
\phi_{q/N_1} \Bigl(\frac{x_1}{z_1},\mu\Bigr) \phi_{\bar{q}/N_2} \Bigl(\frac{x_2}{z_2},\mu\Bigr) , \nonumber
\end{eqnarray}
where the TMD kernel $W(z_1,z_2, \mathbf{q}_{\perp},\mu) $ is defined as
\begin{eqnarray}
W(z_1,z_2, \mathbf{q}_{\perp},\mu) &=& \int  d^2 \mathbf{k}_{1\perp} d^2 \mathbf{k}_{2\perp}
d^2 \bta  \delta^{(2)} ( \mathbf{k}_{1\perp} +\mathbf{k}_{2\perp} -\bta -\mathbf{q}_{\perp}) \nonumber \\
&&\times S(\bta,\mu)
 J_{qq}(z_1, \mathbf{k}_{1\perp},\mu) J_{\bar{q}\bar{q}}(z_2, \mathbf{k}_{2\perp},\mu). 
\end{eqnarray}
All the information on the transverse momentum resides in  $W(z_1,z_2, \mathbf{q}_{\perp}) $, which should be IR finite.
It will be computed at next-to-leading order.

The anomalous dimension for $W$ at one loop is given as
\begin{eqnarray}
\gamma_W &=& \mu \frac{d}{d\mu} W(z_1,z_2, \mathbf{q}_{\perp},\mu)  \nonumber \\
&=& \gamma_S (\mathbf{q}_{\perp}^2) \delta (1-z_1) \delta (1-z_2) 
+\gamma_{J_{qq}} (z_1, \mathbf{q}_{\perp}^2) \delta (1-z_2) +\gamma_{J_{\bar{q}\bar{q}}} (z_2, \mathbf{q}_{\perp}^2)
\delta (1-z_1),
\end{eqnarray}
where
\begin{equation}
\gamma_S = \mu\frac{d}{d\mu} S(\mathbf{q}_{\perp},\mu), \ \ \gamma_{J_{qq}} 
=\mu\frac{d}{d\mu} J_{qq}(z_1, \mathbf{q}_{\perp},\mu), \ \ \gamma_{J_{\bar{q}\bar{q}}} =
\mu\frac{d}{d\mu} J_{\bar{q}\bar{q}} (z_2, \mathbf{q}_{\perp},\mu).
\end{equation}
The anomalous dimensions $\gamma_S$, $\gamma_{J_{qq}}$ and $\gamma_{J_{\bar{q}\bar{q}}}$ contain IR divergences, 
hence not  physically meaningful. But the sum is independent of the IR divergence.

Summarizing our approach, all the particles are on their lightcones, and we define a universal TMDPDF in terms of collinear fields only,
and employ the soft zero-bin subtraction in the TMDPDF to avoid double counting. 
It is different from other approaches in which soft parts are involved in the definition of the TMDPDF.  If soft parts are 
included in the TMDPDF, its definition cannot be universal since soft parts are different in different scattering processes 
in general.

\section{Radiative corrections of the TMDPDF\label{tmdpdf}}
In SCET, the TMDPDF at the parton level is defined as
\begin{equation} \label{tmdpdfd}
f_{q/N} (\frac{\omega}{p_-}, \mathbf{k_{\perp}}) =\langle N (p^-)| \overline{\chi}_n \frac{\FMslash{\overline{n}}}{2}
\delta (\omega -\overline{n}\cdot \mathcal{P}) \delta^{(2)} (\mathbf{k}_{\perp} -\mathbf{\mathcal{P}}_{\perp}) 
\chi_n |N (p^-)\rangle.
\end{equation}
It is implied that $\omega \ge 0$, and the longitudinal momentum fraction $x$ is given by $x=\omega/p^-$. 
The TMDPDF $f_{\overline{q}/N}$ can be treated in a similar way.
As emphasized, it consists of the collinear fields only, which enables a universal definition of the TMDPDF. There is rapidity divergence,
and it is regulated by the $\delta$ regulator in the collinear Wilson lines.  After the zero-bin subtraction, the radiative correction
is free of rapidity divergence as in the full theory.

The integrated PDF can be obtained by integrating the TMDPDF over the transverse momentum to all orders
in $\alpha_s$ as
\begin{equation} \label{pdfntmd}
\phi_{q/N} (x) =\int d^2 \mathbf{k}_{\perp}  f_{q/N} (\frac{\omega}{p_-}, \mathbf{k_{\perp}})  
=\langle N (p^-)| \overline{\chi}_n 
\frac{\FMslash{\overline{n}}}{2}\delta (\omega -\overline{n}\cdot \mathcal{P})  \chi_n |N (p^-)\rangle.
\end{equation}
This is obviously true at tree level since the tree-level PDFs are given by
\begin{equation}
f_{q/N}^{(0)} (x,\kp) =\delta (1-x) \delta^{(2)} (\mathbf{k}_{\perp}), \ \phi_{q/N}^{(0)} (x)= \delta (1-x). 
\end{equation}
But one of the main issues involved in TMDPDF is whether it holds true at higher orders \cite{Ji:2004wu}.  We are going to
construct a formalism in which Eq.~(\ref{pdfntmd}) works to all orders, and show the explicit result at one loop.

The $\delta$ regulator is introduced in the collinear Wilson lines $W_n$ and $W_{\bar{n}}$ as
\begin{equation}
W_n = \sum_{\mathrm{perm}} \exp \Bigl[-\frac{g}{\overline{n}\cdot \mathcal{P} +\delta_1} \overline{n}\cdot A_n\Bigr], 
\ W_{\bar{n}} =  \sum_{\mathrm{perm}} \exp \Bigl[-\frac{g}{n\cdot \mathcal{P} +\delta_2} n\cdot A_{\bar{n}}\Bigr].
\end{equation}
The $\delta$ regulators, as discussed, are employed to regulate the rapidity divergence in TMDPDF. 
These regulators can be either arbitrary parameters independent of kinematics, or  the quantities depending on the details of the 
jets other than the $n$ or $\overline{n}$ directions.  In the special case for a back-to-back current,
if we put $n$ and $\overline{n}$-collinear particles slightly off shell by $p_1^2$ and $p_2^2$, they are given by 
$\delta_1 = p_2^2/n\cdot p_2$ and $\delta_2 =p_1^2/\overline{n}\cdot p_1$ in the latter case. 
\begin{figure}[b] 
\begin{center}
\includegraphics[height=4.0cm]{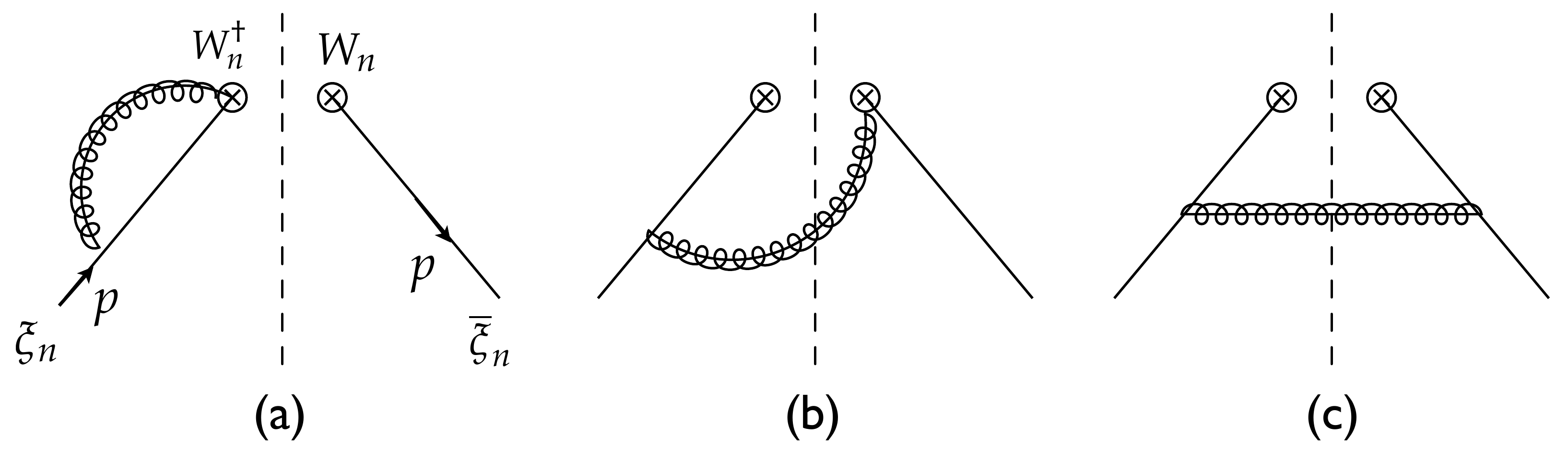}
\end{center}  
\vspace{-0.3cm}
\caption{Feynman diagrams for one-loop corrections of the TMDPDF  (a) virtual corrections and (b), (c) real gluon emission. 
The mirror images of (a) and (b) should be included.\label{feynco}}
\end{figure}

The Feynman diagrams for the radiative corrections at one loop are shown in Fig.~\ref{feynco}. 
Fig.~\ref{feynco} (a) gives
\begin{equation} \label{coa}
M_a = \frac{\alpha_s C_F}{2\pi^2} \delta (1-x) \delta (\mathbf{k}_{\perp}^2) \Bigl[ \frac{1}{\epsilon} (1+\ln \frac{\delta_1}{p^-})
+\ln \frac{\mu^2}{-p^2} -\frac{1}{2} \ln^2 \frac{\delta_1}{p^-} +\ln \frac{\mu^2}{-p^2}  \ln \frac{\delta_1}{p^-} +2 -\frac{\pi^2}{3}
\Bigr].
\end{equation}
Here we use the relation $\delta^{(2)} (\mathbf{k}_{\perp})  = \delta (\mathbf{k}_{\perp}^2)/\pi$ since the remaining function is 
independent of the azimuthal angle.
Fig.~\ref{feynco} (b) and (c) yield
\begin{eqnarray} \label{cob}
M_b &=& \frac{\alpha_s C_F}{2\pi^2} \Bigl\{ \delta (\mathbf{k}_{\perp}^2) \Bigl[ 
  \Bigl( \frac{x}{(1-x)_+} -\delta (1-x) \ln \frac{\delta_1}{p^-}\Bigr) \ln \frac{\mu^2}{-p^2} \nonumber \\
&&+\Bigl( \frac{\pi^2}{6} +
\frac{1}{2} \ln^2 \frac{\delta_1}{p^-}\Bigr) \delta (1-x) 
-x \Bigl( \frac{\ln (1-x)}{(1-x)}\Bigr)_+ -\frac{x \ln x}{(1-x)_+} \Bigr] \nonumber \\
&&+ \Bigl[ \frac{1}{\mathbf{k}_{\perp}^2}\Bigr]_{\mu^2} \Bigl( \frac{x}{(1-x)_+} -\delta (1-x) \ln \frac{\delta_1}{p^-} \Bigr) \Bigr\},
\nonumber \\
M_c &=& \frac{\alpha_s C_F}{2\pi^2} (1-x) \Bigl\{ \delta (\mathbf{k}_{\perp}^2) \Bigl[\ln \frac{\mu^2}{-p^2} 
-1 - \ln x (1-x)\Bigr]  + \Bigl[ \frac{1}{\mathbf{k}_{\perp}^2}\Bigr]_{\mu^2}\Bigr\}.
\end{eqnarray}

In Eq.~(\ref{cob}),  the $\mu^2$ distribution function is introduced. 
For a well-behaved test function $f(\kp)$, the integration involving the $\mu^2$ distribution is  defined as
\begin{equation} \label{mudis}
\int_0^{\Lambda_T^2} d\kp [g(\kp)]_{\mu^2} f(\kp) = \int_0^{\Lambda_T^2} d\kp g(\kp) f(\kp) -
\int_0^{\mu^2} d\kp g(\kp) f(0),
\end{equation}
where $\Lambda_T$ is a finite upper limit for the relevant physical processes in consideration.  
Here $g(\kp)$ is a function which diverges at $\kp=0$. At one loop, there are two types of $g(\kp)$ given by
\begin{equation}
g_1 (\kp) =\frac{1}{\kp},  \ g_2 (\kp) = \frac{\ln (\kp/\mu^2)}{\kp}.
\end{equation}
Of course, any definition of the 
distribution function will do with an arbitrary upper limit in the 
second integral in Eq.~(\ref{mudis}) instead of $\mu^2$ as long as it removes the IR singularity at $\kp=0$. 
The choice of the $\mu^2$ distribution is 
somewhat cosmetic in the sense that  the $\mu$ dependence of the finite part is the same as the radiative corrections of the integrated
PDF.  And there appears no
explicit dependence on $\Lambda_T$.

 Since we require that $\mathbf{k}_{\perp} \sim \mathcal{O} (Q\lambda)$,  the transverse momentum
 should remain finite. It is convenient to obtain the information about the
dependence of the TMDPDF on the small transverse momentum from experiments. 
But in the effective theory, it can reach infinity and the UV 
divergence can be extracted.  However, it becomes subtle if we allow the transverse momentum to have arbitrary 
values extending to infinity.
In this case the distribution should have a different form and the UV divergence should be included. 
This is relevant when we try to obtain the integrated PDF from the TMDPDF by integrating the transverse momentum from 0 to infinity.
Then the definition of the distribution function includes the UV divergence, and it is referred to as the ``infinity distribution". 
This will be discussed in the next section when we compare the radiative corrections of the
TMDPDF and the integrated PDF. The properties of these distribution functions are described in detail in Appendix A.

In computing collinear matrix elements, the kinematic region with soft momentum is also included. Since the soft contribution is computed
separately, the contribution from the soft region to the collinear part should be subtracted to avoid double counting. This is referred to as
the zero-bin subtraction \cite{Manohar:2006nz}.
However, there are two possible kinematic regions for the zero-bin contribution:
soft particles with momentum $p_s \sim Q\lambda$, and usoft particles with momentum 
$p_{\mathrm{us}} \sim Q\lambda^2$. The need for the distinction between the soft and usoft zero-bin contributions arises
depending on whether the processes in consideration are sensitive to soft momentum or not. 
When the zero-bin subtraction was first applied to a heavy-to-light current \cite{Manohar:2006nz}, 
the soft interaction is decoupled from the current because
the interaction puts a heavy or a collinear particle off the mass shell. The resultant effective theory is $\mathrm{SCET}_{\mathrm{I}}$.
In order to go down to $\mathrm{SCET}_{\mathrm{II}}$, where the fields have fluctuations of order $Q\lambda^2$, all the soft scales 
of order $Q\lambda$ are rescaled to usoft scales of order $Q\lambda^2$. Therefore there is no need to distinguish
soft and usoft zero-bin subtractions  because there is a smooth transition from soft to usoft 
contributions as we scale down from $Q\lambda$ to $Q\lambda^2$.

When there is no smooth transition from soft to usoft region as in the Drell-Yan process with transverse momentum of 
order $Q\lambda$,
care should be taken about which zero-bin subtraction should be performed. If we scale down from soft to usoft scales, 
while the transverse
momentum is fixed $q_T \sim Q\lambda$, the resultant usoft interactions cannot produce transverse momentum of order $Q\lambda$. In
other words, the soft part treats physics with momenta of order $Q\lambda$, and a naive computation of the collinear part includes
the soft part which must belong to the soft sector. Therefore the soft zero-bin subtraction is appropriate  to correctly avoid
double counting  in the collinear part.  As we mentioned earlier, the usoft Wilson lines
cancel and there is no usoft contribution.

The zero-bin contribution for $M_a$ is given by
\begin{equation} \label{coa0}
M_a^0 = \frac{\alpha_s C_F}{2\pi^2} \delta (1-x) \delta (\mathbf{k}_{\perp}^2) \Bigl[ -\frac{1}{\epsilon^2} 
-\frac{1}{\epsilon} \ln \frac{\mu^2 p^-}{-p^2 \delta_1} -\frac{1}{2} \ln^2 \frac{ \mu^2 p^-}{-p^2 \delta_1}
-\frac{\pi^2}{4}\Bigr].
\end{equation}
The same result is obtained whether the soft or usoft zero-bin contribution is considered since there is no distinction
in the virtual correction. The major difference comes from real gluon emission.
The soft zero-bin contribution for $M_b$, responsible for real gluon emission, is given as
\begin{eqnarray} \label{cob0}
M_{b,\mathrm{soft}}^0 &=& \frac{\alpha_s C_F}{2\pi^2} \delta (1-x) \Bigl\{\delta (\mathbf{k}_{\perp}^2) 
\Bigl( \frac{1}{2} \ln^2 \frac{\mu^2 p^-}{-p^2\delta_1}  
+\frac{\pi^2}{3} \Bigr) \nonumber \\
&+&\Bigl(\ln \frac{\mu^2}{-p^2} -\ln \frac{\delta_1}{p^-}\Bigr) \Bigl[\frac{1}{\mathbf{k}_{\perp}^2}\Bigr]_{\mu^2}
+\Bigl[\frac{\ln \mathbf{k}_{\perp}^2/\mu^2}{\mathbf{k}_{\perp}^2}\Bigr]_{\mu^2}\Bigr\}.
\end{eqnarray}
On the other hand, the corresponding usoft zero-bin contribution is given by
\begin{equation} \label{busoft}
M_{b,\mathrm{usoft}}^0 = \frac{\alpha_s C_F}{2\pi^2} \delta (1-x) \delta (\mathbf{k}_{\perp}^2)  \Bigl[ \frac{1}{\epsilon^2}
+\frac{1}{\epsilon} \ln \frac{\mu^2 p^-}{-p^2 \delta_1} +\frac{1}{2}\ln^2 \frac{\mu^2 p^-}{-p^2 \delta_1} +\frac{\pi^2}{4}\Bigr].
\end{equation}
This is exactly the same as $M_a^0$ with the opposite sign.
The zero-bin contribution for $M_c$ is suppressed by $\lambda$ compared to $M_c$ and is neglected here. 
Combining Eqs.~(\ref{coa}), (\ref{cob}),  (\ref{coa0}) and  
(\ref{cob0}), the collinear contributions after the soft zero-bin subtraction are given as
\begin{eqnarray}
\tilde{M}_a &=& M_a -M_a^0 = \frac{\alpha_s C_F}{2\pi^2} \delta (1-x) \delta (\mathbf{k}_{\perp}^2) \Bigl[
\frac{1}{\epsilon^2} +\frac{1}{\epsilon} \Bigl(1+\ln \frac{\mu^2}{-p^2}\Bigr) +\ln \frac{\mu^2}{-p^2} +\frac{1}{2}
\ln^2 \frac{\mu^2}{-p^2} +2 -\frac{\pi^2}{12}\Bigr], \nonumber \\
\tilde{M}_b &=& M_b - M_{b,\mathrm{soft}}^0 =\frac{\alpha_s C_F}{2\pi^2} \Bigl\{ \delta(\mathbf{k}_{\perp}^2) \Bigl[
\delta (1-x) \Bigl( -\frac{1}{2}\ln^2  \frac{\mu^2}{-p^2} -\frac{\pi^2}{6}\Bigr) 
+\frac{x}{(1-x)_+} \Bigl(\ln \frac{\mu^2}{-p^2} -\ln x\Bigr)\nonumber \\
&-&x \Bigl( \frac{\ln (1-x)}{1-x} \Bigr)_+ \Bigr] 
+\Bigl[\frac{1}{\mathbf{k}_{\perp}^2}\Bigr]_{\mu^2} \Bigl(\frac{x}{(1-x)_+} - \delta(1-x) \ln  \frac{\mu^2}{-p^2}\Bigr) 
-\delta (1-x)  \Bigl[\frac{1}{\mathbf{k}_{\perp}^2} 
\ln \frac{\mathbf{k}_{\perp}^2}{\mu^2} \Bigr]_{\mu^2} \Bigr\}, \nonumber \\
\tilde{M}_c &=& M_c = \frac{\alpha_s C_F}{2\pi^2} (1-x) \Bigl\{ \delta (\mathbf{k}_{\perp}^2) \Bigl[\ln \frac{\mu^2}{-p^2} 
-1 - \ln x (1-x)\Bigr]  + \Bigl[ \frac{1}{\mathbf{k}_{\perp}^2}\Bigr]_{\mu^2}\Bigr\}. 
\end{eqnarray}
Just for reference,  $\tilde{M}_b$ in the usoft zero-bin subtraction is given as
\begin{eqnarray}
\tilde{M}_{b,\mathrm{usoft}} = M_b - M_{b,\mathrm{usoft}}^0
 &=&\frac{\alpha_s C_F}{2\pi^2} \Bigl\{ \delta(\mathbf{k}_{\perp}^2) \Bigl[
\delta (1-x) \Bigl( -\frac{1}{2}\ln^2  \frac{\mu^2}{-p^2} -\frac{\pi^2}{12}\Bigr) \nonumber \\
&+&\frac{x}{(1-x)_+} \Bigl(\ln \frac{\mu^2}{-p^2} -\ln x\Bigr)-x \Bigl( \frac{\ln (1-x)}{1-x} \Bigr)_+ \Bigr] \nonumber \\
&+&\Bigl[\frac{1}{\mathbf{k}_{\perp}^2}\Bigr]_{\mu^2} \frac{x}{(1-x)_+} - \delta (1-x) \Bigl[ \frac{1}{\mathbf{k}_{\perp}^2}
\Bigr]_{\mu^2} \ln \frac{\delta_1}{p^-}\Bigr\}. 
\end{eqnarray}

Each Feynman diagram for the collinear contribution depends on the $\delta$ regulator, but after the soft zero-bin subtraction, 
the collinear contribution is independent of $\delta_1$. On the other hand, after the
usoft zero-bin subtraction, there still remains the dependence on $\delta_1$. This reflects the fact that there is a mismatch in 
counting the degrees of freedom in the usoft zero-bin contribution and in the soft contribution.   Furthermore, 
note that  $M_a^0+M_{b,\mathrm{usoft}}^0 =0$ from Eqs.~(\ref{coa0})
and (\ref{busoft}),  which confirms that there are no usoft contributions in this process.

To summarize, the soft zero-bin subtraction is appropriate not because it makes the result free of $\delta_1$, but because 
there is a definite physical reason. Since the collinear contribution with the soft zero-bin subtraction is free of the $\delta$ regulator,
there is no need to relate the $\delta$ regulators in the collinear Wilson lines to the offshellness appearing in the soft Wilson lines. 
This is important because the treatment of the
$\delta$ regulator can be applied to processes other than Drell-Yan process, in which there are collinear particles other than 
$n$ and $\overline{n}$ directions. The collinear Wilson line is obtained by integrating out the degrees 
of freedom of order $Q$ when collinear gluons are emitted from other particles. 
For example, if there are  distinct jets or heavy colored particles in the final state, 
the $\delta$ regulator for a specific collinear Wilson line 
depends, in general, on the offshellness of other  jets.
The leading eikonalized form is the same irrespective of the source of the emitted collinear gluons. If we try to incorporate
the $\delta$ regulator by invoking the offshellness of other particles, the $\delta$ regulator depends in a complicated way on
the offshellness of other particles. 
In our scheme, it is not necessary for the $\delta$ regulator in the collinear Wilson line to have a definite relationship with the  
offshellness in the soft part. The dependence on the $\delta$ regulator in the collinear part disappears when appropriate 
zero-bin contributions are subtracted, not from the soft part.

In fact, we do not have to relate the $\delta$ regulators to the offshellness of the particles. The $\delta$ 
regulators may be used also in the soft Wilson lines, but we stress that the $\delta$ regulators in the collinear and soft Wilson lines
are independent regulators. They can be related only when we obtain the $\delta$ regulators from the offshellness in the back-to-back
current. Otherwise, the attempt to assign some relations between them may simplify the computation, but it evades the consistent treatment
of the rapidity divergence.

The total collinear contribution including the self-energy corrections with the soft zero-bin subtraction is given at one loop as
\begin{eqnarray} \label{tmdpdfco}
M_{\mathrm{col}} &=& 2(M_a -M_a^0) + 2(M_b -M_{b,\mathrm{soft}}^0) +M_c  + (Z_{\xi} + R_{\xi}) 
\delta (1-x) \frac{\delta (\mathbf{k}_{\perp}^2)}{\pi} \nonumber \\
&=& \frac{\alpha_s C_F}{2\pi^2} \Bigl\{ \delta (\mathbf{k}_{\perp}^2)
\Bigl[ \delta(1-x) \Bigl( \frac{2}{\epsilon^2} +\frac{1}{\epsilon} (\frac{3}{2} +2 \ln \frac{\mu^2}{-p^2})
+\frac{7}{2} -\frac{\pi^2}{2} +\frac{3}{2} \ln \frac{\mu^2}{-p^2}\Bigr)\nonumber \\
&&+\frac{1+x^2}{(1-x)_+} \Bigl( \ln \frac{\mu^2}{-p^2} -\ln x\Bigr) -(1+x^2)\frac{\ln (1-x)}{(1-x)_+} \Bigr]
-(1-x)\nonumber \\
&& +\Bigl(\frac{1+x^2}{(1-x)_+} -2\delta (1-x) \ln \frac{\mu^2}{-p^2}\Bigr) \Bigl[\frac{1}{\mathbf{k}_{\perp}^2}\Bigr]_{\mu^2}  
-2 \delta (1-x) \Bigl[\frac{1}{\mathbf{k}_{\perp}^2} \ln \frac{\mathbf{k}_{\perp}^2}{\mu^2}\Bigr]_{\mu^2} \Bigr\},
\end{eqnarray}
where $Z_{\xi}$ and $R_{\xi}$ are the wave function renormalization  and the residue for the collinear fermion $\xi$ 
at order $\alpha_s$. They are given by
\begin{equation}
Z_{\xi} =-\frac{\alpha_s C_F}{4\pi}\frac{1}{\epsilon}, \ R_{\xi} = -\frac{\alpha_s C_F}{4\pi}\Bigl(1+
\ln \frac{\mu^2}{-p^2}\Bigr).
\end{equation}
The collinear contribution, or the TMDPDF is independent of $\delta$. That is, it is free of rapidity divergence.
It is also true in the radiative corrections for the TMDPDF $f_{\overline{q}/N} (x,\kp)$. Note that the mixing
between the UV and IR divergences still remains. The mixing is cancelled if the soft part is added.

\section{Soft function and its one-loop corrections\label{softone}}
The soft function is defined as
\begin{equation}
S(\bta) =\frac{1}{N_c} \langle 0|\mathrm{tr} \Bigl[ Y_n^{\dagger} Y_{\overline{n}} \delta^{(2)}
(\bta +i\nabla_{\perp}) Y_{\overline{n}}^{\dagger} Y_n  \Bigr] |0\rangle.
\end{equation}
The soft Wilson lines with the offshellness to regulate IR divergence are given by
\begin{equation}
Y_n = \sum_{\mathrm{perm}} \exp \Bigl[ -\frac{1}{n\cdot \mathcal{P}+\Delta_1 +i0} gn\cdot A_s\Bigr], \ Y_{\overline{n}} =
\sum_{\mathrm{perm}} \exp \Bigl[ -\frac{1}{\overline{n}\cdot \mathcal{P}+\Delta_2 +i0} g\overline{n}\cdot A_s\Bigr].
\end{equation}
The operators $n\cdot \mathcal{P}$ and $\overline{n}\cdot \mathcal{P}$ extract the $n$- and $\overline{n}$-components 
of the momenta from the soft gluon.
The appearance of $\Delta_1$ and $\Delta_2$ in $Y_n$ and $Y_{\overline{n}}$ resembles the $\delta$ regulators in the collinear 
Wilson lines, but they are from different origins. Recall that the soft Wilson lines are obtained by integrating out the 
offshellness of the intermediate states when soft 
gluons are emitted from a collinear particle. Therefore, unlike the collinear Wilson lines, $\Delta_i$'s depend solely on the offshellness of
the relevant collinear particle, not other jets. 
And it should be noted that these regulators in the soft Wilson lines regulate the IR divergence, not the rapidity
divergence.  We can write $\Delta_1 = p_1^2/p_1^-$, $\Delta_2= p_2^2/p_2^+$. Only
in the case of the back-to-back current, the $\delta$ regulators in the collinear Wilson line are given by  $\Delta_1 =\delta_2$ and $\Delta_2=\delta_1$.

\begin{figure}[t] 
\begin{center}
\includegraphics[height=4.0cm]{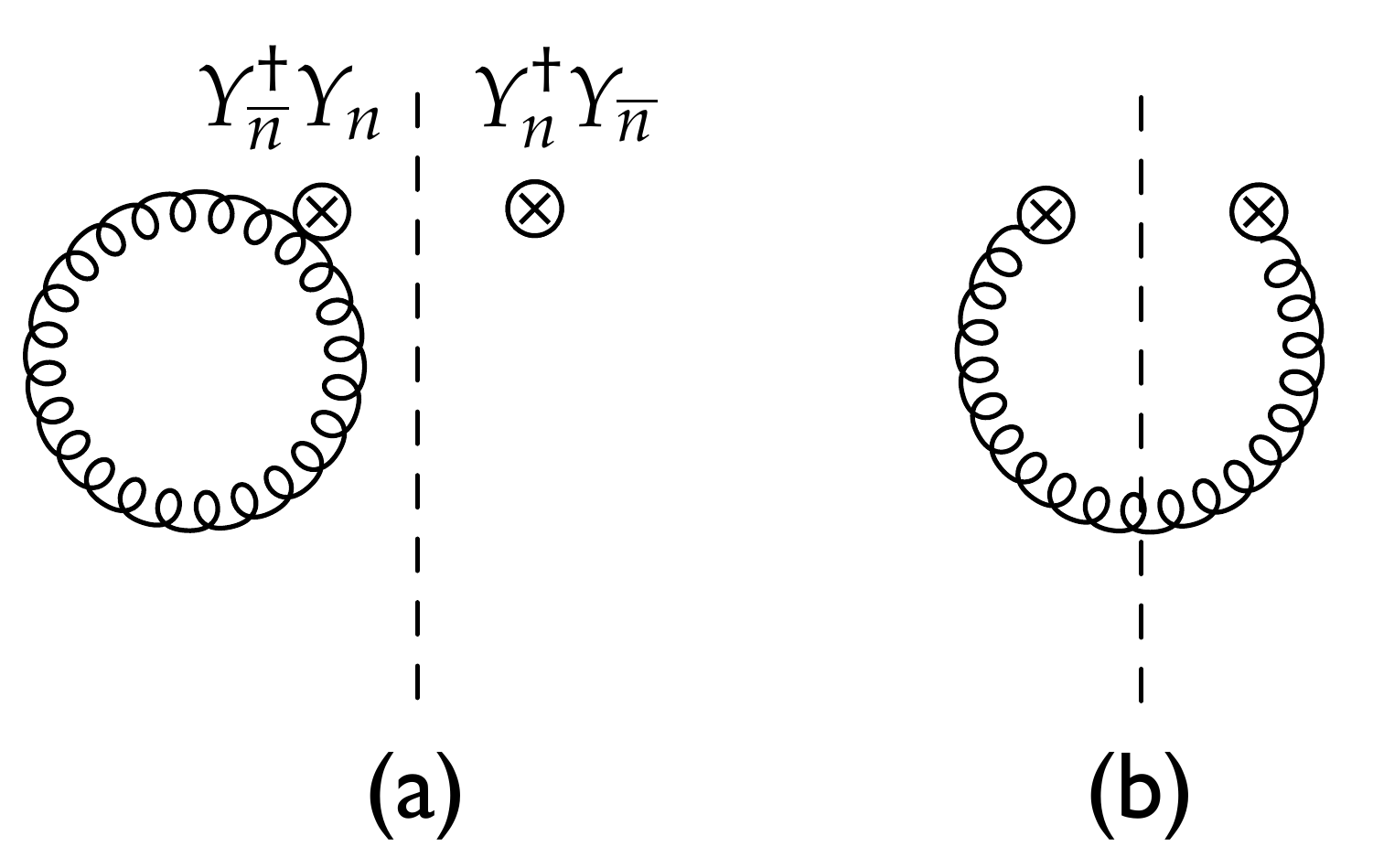}
\end{center}  
\vspace{-0.3cm}
\caption{Feynman diagrams for soft contributions (a) virtual corrections and (b) real gluon emission.\label{feynsoft}}
\end{figure}

The Feynman diagrams for the radiative correction of $S(\bta)$ at one loop are shown in Fig.~\ref{feynsoft}. The virtual correction
in Fig.~\ref{feynsoft} (a), and the real gluon emission in Fig.~\ref{feynsoft} (b) with their hermitian conjugates are given by
\begin{eqnarray}
M_a^s &=& \frac{\alpha_s C_F}{2\pi^2} \delta (\bta^2) \Bigl( -\frac{2}{\epsilon^2} -\frac{2}{\epsilon} \ln \frac{\mu^2}{\Delta_1
\Delta_2}  -\ln^2 \frac{\mu^2}{\Delta_1 \Delta_2} +\frac{\pi^2}{2} \Bigr), \nonumber \\
M_b^s &=& \frac{\alpha_s C_F}{2\pi^2} \Bigl[ \delta (\bta^2) \Bigl( \ln^2 \frac{\mu^2}{\Delta_1 \Delta_2} -\frac{\pi^2}{3} \Bigr) +2\Bigl[\frac{1}{\bta^2} 
\ln \frac{\bta^2}{\Delta_1 \Delta_2}\Bigr]_{\mu^2}\Bigr], 
\end{eqnarray}
and the total soft contribution is given by
\begin{equation} \label{tmdpdfso}
M_{\mathrm{soft}} = M_a^s + M_b^s = \frac{\alpha_s C_F}{2\pi^2} \Bigl[ \delta (\bta^2) \Bigl( -\frac{2}{\epsilon^2} 
-\frac{2}{\epsilon} \ln \frac{\mu^2}{\Delta_1 \Delta_2} +\frac{\pi^2}{6}\Bigr) +\Bigl[\frac{2}{\bta^2} 
\ln \frac{\bta^2}{\Delta_1 \Delta_2}\Bigr]_{\mu^2}\Bigr].
\end{equation}
The soft function depends on the IR regulators $\Delta_i $.  Especially, there is mixing between the UV and IR divergences. 
When it is combined with the collinear contributions 
in Eq.~(\ref{tmdpdfco}) with $p^2=p_1^2$, and another collinear contribution from the TMDPDF in the $\overline{n}$ direction
with $p^2 =p_2^2$,  the mixing of the UV and IR divergences cancels out.

\section{TMDPDF and integrated PDF\label{comp}}

The integrated PDF is defined also in terms of the collinear fields only as
\begin{equation} \label{pdfd}
\phi_{q/N} (\frac{\omega}{p_-}) =\langle N (p^-)| \overline{\chi}_n \frac{\FMslash{\overline{n}}}{2}
\delta (\omega -\overline{n}\cdot \mathcal{P}) 
\chi_n |N (p^-)\rangle.
\end{equation}
Compared to the definition of the TMDPDF, the only difference is the absence of the two-dimensional delta function on the transverse
momentum. That is, the integrated PDF can be obtained from TMDPDF by integrating over the transverse momentum. It looks trivial
to obtain the relation between the TMDPDF and the integrated PDF, but it is not trivial to verify the relation 
at higher orders \cite{Ji:2004wu}. The complication
arises due to the regularization methods employed to tame the UV, IR and rapidity divergences. 
The UV divergence can be typically treated in
dimensional regularization. The rapidity divergence is handled by the $\delta$ regulator in the collinear Wilson line with the soft
zero-bin subtraction. The IR divergence is taken care of only when the collinear and soft parts are summed. In this procedure, the collinear
and soft parts are entangled because the regularization methods complicate the transparent relation  between
the TMDPDF and the integrated PDF, and it needs special care to see the relation at higher orders.

The important issue in establishing the relation is the range of the transverse momentum. When the transverse momentum is to be 
integrated, the region of integration should be determined beforehand. If we confine ourselves to finite transverse momentum, there is
no UV divergence in real gluon emissions.  Since the Drell-Yan process with the transverse momentum $q_T \sim Q\lambda$ is considered,
we can limit the region of validity accordingly. In that case, we use the idea of $\mu^2$ distribution which is suggested  exactly for 
this purpose. On the other hand, the transverse momentum can extend to infinity in the effective theory, 
and the $\mu^2$ distribution is not valid any more. There are two possible ways  to establish the
relation between the TMDPDF and the integrated PDF. First, the radiative corrections for the integrated PDF are performed in 
the usual way, and we modify the $\mu^2$ distribution in the TMDPDF to include the infinite transverse momentum. 
Secondly, in the spirit of computing TMDPDF assuming that the transverse momentum
never reaches infinity but some finite scale $\Lambda_T$, the radiative corrections of the integrated PDF are modified such that
the integration over the transverse momentum is limited to $\Lambda_T$ in real gluon emissions, 
while the virtual corrections are performed as usual.

Let us first consider the case in which the transverse momentum can go to infinity.  The radiative corrections to the integrated PDF can be 
computed using the Feynman diagrams in Fig.~\ref{feynco}   without the two-dimensional delta function for the 
transverse momentum. We organize the collinear matrix elements for integrated PDF including the zero-bin subtractions in terms 
of the virtual and real gluon emissions as
\begin{eqnarray}
M_{\mathrm{PDF}}^V &=& 2\tilde{M}_a +(Z_{\xi} +R_{\xi}) \delta (1-x) \frac{\delta (\kp)}{\pi} \nonumber \\
&=& \frac{\alpha_s C_F}{2\pi} \delta (1-x) \Bigl( \frac{2}{\epsilon^2} +\frac{3}{2\epsilon} +\frac{2}{\epsilon} 
\ln \frac{\mu^2}{-p^2} +\frac{3}{2} \ln \frac{\mu^2}{-p^2} +\ln^2 \frac{\mu^2}{-p^2} +\frac{7}{2} -\frac{\pi^2}{6} \Bigr),
\nonumber \\
M_{\mathrm{PDF}}^R &=& 2\tilde{M}_b +M_c \nonumber \\
&=&\frac{\alpha_s C_F}{2\pi} \Bigl[ \delta (1-x) \Bigl( -\frac{2}{\epsilon^2} -\frac{2}{\epsilon} 
\ln \frac{\mu^2}{-p^2} -\ln^2 \frac{\mu^2}{-p^2} -\frac{\pi^2}{6}\Bigr)    \nonumber \\
&+& \frac{1+x^2}{(1-x)_+} \Bigl( \frac{1}{\epsilon} +\ln \frac{\mu^2}{-p^2} -\ln x\Bigr)  - (1+x^2) 
\Bigl( \frac{\ln (1-x)}{1-x}\Bigr)_+-2(1-x)\Bigr].
 \end{eqnarray}
And the total collinear contributions at one loop are given as
\begin{eqnarray} \label{pdfco}
M_{\mathrm{PDF}}&=& M_{\mathrm{PDF}}^V +M_{\mathrm{PDF}}^R  \nonumber \\
&=& \frac{\alpha_s C_F}{2\pi}  \Bigl[ \delta (1-x) \Bigl( \frac{3}{2\epsilon} +\frac{3}{2} 
\ln \frac{\mu^2}{-p^2}+\frac{7}{2} -\frac{\pi^2}{3}  \Bigr) -2 (1-x) \nonumber \\
&&+\frac{1+x^2}{(1-x)_+} \Bigl( \frac{1}{\epsilon} + \ln \frac{\mu^2}{-p^2} -\ln x\Bigr) - (1+x^2) \Bigl(\frac{\ln (1-x)}{1-x}\Bigr)_+ \Bigr].
\end{eqnarray}
The result is independent of  the $\delta$ regulators, which means that there is no rapidity divergence in the integrated PDF.

For the TMDPDF, we need another distribution called
the ``infinity" distribution, in which the finite scale $\Lambda_T$ in defining the $\mu^2$ distribution extends to infinity. 
The function of the form
\begin{equation}
g_1(\kp, \epsilon,\Delta) = \frac{\mu^{2\epsilon}}{(\kp +\Delta)^{1+\epsilon}},
\end{equation}
appears in the one-loop correction, where $\Delta$ is a small quantity.  For a regular 
test function $f(\kp)$, the integral of $f(\kp)$ multiplied by $g_1 (\kp, \epsilon,\Delta)$ is written as
\begin{eqnarray}
\int_0^{\infty} d\kp g_1 (\kp,\epsilon,\Delta) f(\kp) &=& \int_0^{\infty} d\kp g_1 (\kp,\epsilon,0) \Bigl( f(\kp)-f(0)\Bigr)
+f(0) \int_0^{\infty} d\kp g_1(\kp,\epsilon,\Delta) \nonumber \\
&=& \int_0^{\infty} d\kp \Bigl[g_1 (\kp, \epsilon,0)\Bigr]_{\infty} f(\kp) + f(0) \Bigl(\frac{1}{\epsilon} +\ln \frac{\mu^2}{\Delta}\Bigr). 
\end{eqnarray}
In the infinity distribution function, we put $\Delta =0$ since the IR divergence is already regulated. 
Compared to the $\mu^2$  distribution function, the only difference is the appearance of  the pole $1/\epsilon$. 
This comes from the UV region, which cannot be reached in the $\mu^2$ distribution function. 
Therefore, $g_1 (\kp, \epsilon,\Delta)$ can be written as
\begin{equation}
g_1 (\kp,\epsilon,\Delta) = \Bigl(\frac{1}{\epsilon} +\ln \frac{\mu^2}{\Delta}\Bigr) \delta (\kp) +\Bigl[g_1 (\kp,\epsilon,0)
\Bigr]_{\infty}.
\end{equation}
Similarly, there is another function expressed in terms of the infinity distribution functions as
\begin{eqnarray}
g_2 (\kp, \epsilon, \Delta) &=& \frac{\mu^{2\epsilon} \displaystyle  \ln  \frac{\kp}{\Delta}}{(\kp -\Delta)^{1+\eps}} \nonumber \\
&=& 
\Bigl( \frac{1}{\epsilon^2} +\frac{1}{\epsilon}\ln \frac{\mu^2}{\Delta} +\frac{1}{2}\ln^2 \frac{\mu^2}{\Delta} +\frac{\pi^2}{3}\Bigr)
\delta (\kp) +\Bigl[\frac{1}{\kp}\Bigr]_{\infty} \ln \frac{\mu^2}{\Delta} +\Bigl[\frac{\ln \kp/\mu^2}{\kp}\Bigr]_{\infty}.
\end{eqnarray}
Here two points should be noted. First, the terms proportional to $\delta (\kp) $ in the $\mu^2$ distribution have the same 
$\mu$ dependence  as those in the infinity distribution except the UV poles. And secondly, the infinity distribution is not 
exactly a distribution function in a rigorous sense. It is because the UV pole cannot be peaked near $\kp \sim 0$, 
therefore it should be understood that the infinity distribution function is meaningful only after the integral over 
$\kp$ is performed. This awkward situation occurs since we are going to compare the 
TMDPDF and the integrated PDF, and the integrated PDF is obtained by integrating $\kp$ over all values including infinity. 

The total radiative correction for the TMDPDF is given as
\begin{eqnarray}
M_{\mathrm{TMDPDF}} &=& 
\frac{\alpha_s C_F}{2\pi^2} \Bigl\{ \delta (\kp) \Bigl[ \delta (1-x) \Bigl( \frac{3}{2\epsilon} +\frac{7}{2} -\frac{\pi^2}{3}
+\frac{3}{2}\ln \frac{\mu^2}{-p^2}\Bigr) -2 (1-x) \nonumber \\
&&+\frac{1+x^2}{(1-x)_+} \Bigl(\frac{1}{\epsilon} +\ln \frac{\mu^2}{-p^2} -\ln x\Bigr) -(1+x^2) \Bigl( \frac{\ln (1-x)}{1-x}\Bigr)_+
\Bigr] \nonumber \\
&&+\frac{1+x^2}{(1-x)_+}\Bigl[\frac{1}{\kp}\Bigr]_{\infty} -2\delta (1-x) \Bigl[\frac{1}{\kp} \ln \frac{\kp}{-p^2}\Bigr]_{\infty} \Bigr\}. 
\end{eqnarray}
If we integrate over $\mathbf{k}_{\perp}$, the infinity distributions vanish and we obtain
\begin{eqnarray}
\int d^2 \mathbf{k}_{\perp} M_{\mathrm{TMDPDF}} &=& 
\frac{\alpha_s C_F}{2\pi} \Bigl[  \delta (1-x) \Bigl( \frac{3}{2\epsilon} +\frac{7}{2} -\frac{\pi^2}{3}
+\frac{3}{2}\ln \frac{\mu^2}{-p^2}\Bigr) -2 (1-x) \nonumber \\
&&+\frac{1+x^2}{(1-x)_+} \Bigl(\frac{1}{\epsilon} +\ln \frac{\mu^2}{-p^2} -\ln x\Bigr) -(1+x^2) \Bigl( \frac{\ln (1-x)}{1-x}\Bigr)_+
\Bigr] ,
\end{eqnarray}
which is exactly the same as the radiative corrections for the integrated PDF in Eq.~(\ref{pdfco}). 

Secondly, we  consider the case in which the transverse momentum has the range $0\le \kp \le \Lambda_T^2$.
Then the radiative corrections for the integrated PDF is handled differently. 
The loop momentum $l$ is defined in $D$ dimensions, and there is a two-dimensional 
delta function. We introduce the momentum vector in $D-4 =-2\epsilon$ dimensions, $\hat{l}_{\perp}$, and 
$\mathbf{l}_{\perp}$ is a two-dimensional momentum vector. Then $l^2$ can be written as 
$l^2 = l^+ l^-  -\mathbf{l}_{\perp}^2 -\hat{l}_{\perp}^2$.
First the two-dimensional transverse momentum 
is fixed in real gluon emissions and the integral over $\hat{l}_{\perp}$  in $D-4$ dimensions is performed first. 
Then  we integrate  over the two-dimensional 
transverse momentum. Since the virtual corrections are not affected, $\tilde{M}_a$ is the same. And the matrix elements
for the real gluon emissions are modified to be
\begin{eqnarray}
M_{\mathrm{PDF}}^R &=& \frac{\alpha_s C_F}{2\pi} \Bigl[\delta (1-x) \Bigl( -\frac{\pi^2}{6} 
-\frac{1}{2} \ln^2 \frac{\Lambda_T^2}{-p^2}\Bigr)
+\frac{1+x^2}{(1-x)_+} \Bigl(\ln \frac{\Lambda_T^2}{-p^2} -\ln x\Bigr) \nonumber \\
&&-(1+x^2)\Bigl( \frac{ \ln (1-x)}{1-x}\Bigr)_+ -(1-x)\Bigr].
\end{eqnarray}
The whole collinear part of the integrated PDF at one loop is given by 
\begin{eqnarray} \label{mupdf}
M_{\mathrm{PDF}} 
&=& \frac{\alpha_s C_F}{2\pi} \Bigl\{ 
\delta(1-x) \Bigl[ \frac{2}{\epsilon^2} +\frac{1}{\epsilon} \Bigl(\frac{3}{2} +2 \ln \frac{\mu^2}{-p^2}\Bigr)
+\frac{7}{2} -\frac{\pi^2}{2} +\frac{3}{2} \ln \frac{\mu^2}{-p^2}+\ln^2 \frac{\mu^2}{-p^2}
-\ln^2 \frac{\Lambda_T^2}{-p^2}\Bigr]\nonumber \\
&&+\frac{1+x^2}{(1-x)_+} \Bigl( \ln \frac{\Lambda_T^2}{-p^2} -\ln x\Bigr) -(1+x^2)\frac{\ln (1-x)}{(1-x)_+} 
-(1-x) \Bigr\}.
\end{eqnarray}
If we integrate Eq.~(\ref{tmdpdfco}) from 0 to $\Lambda_T^2$, Eq.~(\ref{mupdf}) is obtained. In either case, the integrated PDF
is consistently obtained by integrating the TMDPDF over the transverse momentum to one loop. 

\section{Differential scattering cross section at NLO\label{renor}}
The TMDPDF can be expressed in terms of the convolution of the integrated PDF and the TMD kernel which contains all the information
on the transverse momentum in the TMDPDF. Its relation, Eq.~(\ref{tmdint}), is given again as 
\begin{equation} \label{conv}
f_{q/N} (x, \kperp, \mu) = \int_x^1 \frac{dz}{z} J_{qq}(z, \kperp,\mu) \phi_{q/N} \Bigl(\frac{x}{z},\mu\Bigr) =
\int_x^1 \frac{dz}{z} J_{qq}\Bigl(\frac{x}{z}, \kperp,\mu \Bigr) \phi_{q/N} (z,\mu),
\end{equation}
where $\phi_{q/N} (z)$ is the integrated PDF.  We will use this relation to obtain the renormalization group behavior 
of the scattering cross section to one loop. Eq.~(\ref{conv}) can be regarded as matching between two SCET's, 
where the TMDPDF is the quantity in which
$p \sim Q\lambda$ and the integrated PDF is the one where $p \sim Q\lambda^2$. And the kernel $J_{qq}(z,\kperp,\mu)$ is 
the Wilson coefficient in matching. 
At zeroth order in $\alpha_s$, $J_{qq}^{(0)}$ and $\phi_{q/N}^{(0)}$ are given by
\begin{equation}
J^{(0)}_{qq} (z,\kperp) =\frac{1}{\pi} \delta (1-z) \delta (\kp), \ \phi_{q/N}^{(0)} (z) =\delta (1-z).
\end{equation}
The TMDPDF and the soft functions at order $\alpha_s$ are obtained from Eqs.~(\ref{tmdpdfco}) and (\ref{tmdpdfso}) by removing the UV
divergent terms. They seem to have the mixed divergent terms, but they disappear when all the contributions are added, therefore it is 
of no concern here. They are given as
\begin{eqnarray}
f_{q/N}^{(1)} (x,\kperp\mu) &=& \frac{\alpha_s C_F}{2\pi^2} \Bigl\{ \delta (\kp) \Bigl[ \delta (1-x) \Bigl( \frac{3}{2} \ip +\frac{7}{2}
-\frac{\pi^2}{2} \Bigr) \nonumber \\
&&+ \frac{1+x^2}{(1-x)_+} \Bigl(\ip -\ln x\Bigr) -(1+x^2) \Bigl(\frac{\ln (1-x)}{1-x})\Bigr)_+ - (1-x) \Bigr] \nonumber \\
&&-2 \delta (1-x) \Bigl[\frac{1}{\kp} \ln \frac{\kp}{-p^2} \Bigr]_{\mu^2} + \frac{1+x^2}{(1-x)_+} \Bigl[\frac{1}{\kp}\Bigr]_{\mu^2}
\Bigr\}, \nonumber \\
S^{(1)} (\bta) &=& \frac{\alpha_s C_F}{2\pi^2} \Bigl\{\frac{\pi^2}{6} \delta (\bta^2) +2\Bigl[\frac{1}{\bta^2} 
\ln \frac{\bta^2}{\Delta_1 \Delta_2}\Bigr]_{\mu^2}\Bigr\}.
\end{eqnarray}

Expanding Eq.~(\ref{conv}) to first order in $\alpha_s$, we obtain
\begin{eqnarray}\label{fqn}
f_{q/N}^{(1)} (x,\kperp) &=& \int_x^1 \frac{dz}{z} \Bigl[J^{(1)} (z,\kperp) \phi_{q/N}^{(0)} (x/z) 
+J^{(0)} (x/z,\kperp) \phi_{q/N}^{(1)} (z) \Bigr]  \nonumber \\
&=& J^{(1)} (x,\kperp) +\frac{1}{\pi} \delta (\kp) \phi_{q/N}^{(1)} (x). 
\end{eqnarray}
The integrated PDF at order $\alpha_s$ is given by
\begin{eqnarray}\label{phiqn}
\phi_{q/N}^{(1)} (x)  &=& \frac{\alpha_s C_F}{2\pi} \Bigl\{ \delta (1-x) \Bigl( \frac{7}{2} -\frac{\pi^2}{3} +\frac{3}{2} \ip \Bigr)
- 2(1-x) \nonumber \\
&&+\frac{1+x^2}{(1-x)_+} \Bigl(\ip -\ln x\Bigr) - (1+x^2) \Bigl(\frac{\ln (1-x)}{1-x}\Bigr)_+ \Bigr\},
\end{eqnarray}
and the anomalous dimension for the PDF is given by
\begin{equation}
\gamma_{\phi} (x)= \frac{\alpha_s }{2\pi} P_{qq} (x)=\frac{\alpha_s C_F}{2\pi} 
\Bigl[\frac{3}{2}\delta (1-z) +\frac{1+z^2}{(1-z)_+}\Bigr].
\end{equation}
Using Eqs.~(\ref{fqn}) and (\ref{phiqn}), the new result is the TMD kernel at order $\alpha_s$, which is given as
\begin{eqnarray}
J^{(1)} (x,\kperp,\mu) &=& f_{q/N}^{(1)} (x,\kperp,\mu) -\frac{1}{\pi}\delta (\kp) \phi_{q/N}^{(1)} (x, \mu)  \\
&=& \frac{\alpha_s C_F}{2\pi^2} \Bigl\{ \delta (\kp) \Bigl(1-x-\frac{\pi^2}{6}\Bigr) +
\frac{1+x^2}{(1-x)_+} \Bigl[  \frac{1}{\kp}\Bigr]_{\mu^2} -2\delta (1-x) \Bigl[\frac{1}{\kp}
\ln \frac{\kp}{-p^2} \Bigr]_{\mu^2} \Bigr\}, \nonumber 
\end{eqnarray}
where $[f(\kp)]_{\mu^2}$ is defined as
\begin{equation}
[f(\kp)]_{\mu^2} = f(\kp) -\delta (\kp) \int_0^{\mu^2} d\mathbf{l}_{\perp}^2 f(\mathbf{l}_{\perp}^2).
\end{equation}

The factorized differential scattering cross section is written as
\begin{eqnarray} \label{finxsec}
\frac{d\sigma}{d^2\mathbf{q}_{\perp}} &=& \int dx_1 dx_2 \sigma_0 H(Q,\mu) 
\int d^2 \mathbf{l}_{1\perp} d^2 \mathbf{l}_{2\perp}
d^2 \bta \delta^{(2)} (\mathbf{q}_{\perp} -\mathbf{l}_{1\perp} -\mathbf{l}_{2\perp} -\bta) S(\bta) \\
&\times& \int_{x_1}^1 \frac{dz_1}{z_1} J_{qq} (z_1,\mathbf{l}_{1\perp}) \phi_{q/N_1} \Bigl(\frac{x_1}{z_1}\Bigr) 
\int_{x_2}^1 \frac{dz_2}{z_2} J_{\bar{q}\bar{q}} (z_2,\mathbf{l}_{2\perp}) \phi_{\bar{q}/N_2} \Bigl(\frac{x_2}{z_2}\Bigr) 
\nonumber \\
&=& \int dx_1 dx_2\sigma_0 H(Q,\mu)    \int_{x_1}^1 \frac{dz_1}{z_1}
\int_{x_2}^1 \frac{dz_2}{z_2} W(z_1,z_2,\mathbf{q}_{\perp},\mu) \phi_{q/N_1} \Bigl(\frac{x_1}{z_1},\mu\Bigr)  
 \phi_{\bar{q}/N_2} \Bigl(\frac{x_2}{z_2},\mu\Bigr) .\nonumber 
\end{eqnarray}
To NLO, the TMD kernel $W(z_1,z_2,\mathbf{q}_{\perp},\mu)$ can be written as
\begin{eqnarray}
W(z_1,z_2,\mathbf{q}_{\perp}) &=& \int d^2 \mathbf{l}_{1\perp} d^2 \mathbf{l}_{2\perp}
d^2 \bta \delta^{(2)} (\mathbf{q}_{\perp} -\mathbf{l}_{1\perp} -\mathbf{l}_{2\perp} -\bta)  S(\bta) 
J_{qq} (z_1,\mathbf{l}_{1\perp})
J_{\bar{q}\bar{q}} (z_2,\mathbf{l}_{2\perp})  \nonumber \\
&=&\int d^2 \mathbf{l}_{1\perp} d^2 \mathbf{l}_{2\perp}
d^2 \bta \delta^{(2)} (\mathbf{q}_{\perp} -\mathbf{l}_{1\perp} -\mathbf{l}_{2\perp} -\bta)  \nonumber \\
&&\times \Bigl[ S^{(1)} (\bta) J_{qq}^{(0)} (z_1,\mathbf{l}_{1\perp}) J_{\bar{q}\bar{q}}^{(0)} (z_2,\mathbf{l}_{2\perp}) 
+S^{(0)} (\bta) J_{qq}^{(1)} (z_1,\mathbf{l}_{1\perp}) J_{\bar{q}\bar{q}}^{(0)} (z_2,\mathbf{l}_{2\perp}) \nonumber \\
&&+S^{(0)} (\bta) J_{qq}^{(0)} (z_1,\mathbf{l}_{1\perp}) J_{\bar{q}\bar{q}}^{(1)} (z_2,\mathbf{l}_{2\perp})  \Bigr] \nonumber \\
&=&\frac{\alpha_s C_F}{2\pi^2} \Bigl\{ \delta (\mathbf{q}_{\perp}^2) \Bigl[\delta (1-z_1) (1-z_2) +\delta (1-z_2) (1-z_1)  
-\frac{\pi^2}{6} \delta (1-z_1) \delta (1-z_2) \Bigr]\nonumber \\
&&+\Bigl[\frac{2}{\mathbf{q}_{\perp}^2} \ln \frac{Q^2}{\mathbf{q}_{\perp}^2}
\Bigr]_{\mu^2} \delta (1-z_1) \delta (1-z_2)\nonumber \\
&&+\Bigl[\frac{1}{\mathbf{q}_{\perp}^2}\Bigr]_{\mu^2} \Bigl[ \frac{1+z_1^2}{(1-z_1)_+} \delta (1-z_2) +\delta (1-z_1)
\frac{1+z_2^2}{(1-z_2)_+} \Bigr] \Bigr\}.
\end{eqnarray}
The anomalous dimension of $W$ at one loop is given as
\begin{eqnarray}
\gamma_W (z_1, z_2, \mathbf{q}_{\perp}^2, \mu) &=& \frac{d}{d\ln \mu} 
W(z_1, z_2,\mathbf{q}_{\perp},\mu) = 2\mu^2 \frac{d}{d\mu^2} W(z_1, z_2,\mathbf{q}_{\perp},\mu)  \\
&=& \frac{\alpha_s C_F}{2\pi} \frac{\delta (\mathbf{q}_{\perp}^2)}{\pi}
 \Bigl[ 2\ln \frac{\mu^2}{Q^2} \delta (1-z_1) \delta (1-z_2)
\nonumber \\
&&-\frac{1+z_1^2}{(1-z_1)_+} \delta (1-z_2)  -\delta (1-z_1) \frac{1+z_2^2}{(1-z_2)_+} \Bigr] \nonumber \\
&=& \frac{\alpha_s}{2\pi} \frac{\delta (\mathbf{q}_{\perp}^2)}{\pi} \Bigl[C_F \Bigl(2\ln \frac{\mu^2}{Q^2} +3 \Bigr) 
 \delta (1-z_1) \delta (1-z_2) \nonumber \\
&&-P_{qq} (z_1) \delta (1-z_2) -P_{\bar{q}\bar{q}} (z_2) \delta (1-z_1)\Bigr], \nonumber 
\end{eqnarray}
where $P_{qq} (z)=P_{\bar{q}\bar{q}} (z)$. It can be also written as
\begin{eqnarray}
\gamma_W (z_1, z_2, \mathbf{q}_{\perp}^2, \mu) &=& \delta (\mathbf{q}_{\perp}^2) \tilde{\gamma}_W (z_1, z_2, \mu) \\
&=& \gamma_{J_{qq}} (z_1, \mathbf{q}_{\perp}^2,\mu) \delta (1-z_1) +\gamma_{J_{\bar{q}\bar{q}}} (z_2, \mathbf{q}_{\perp}^2,\mu) 
\delta (1-z_2) \nonumber \\
&&+\gamma_S (\mathbf{q}_{\perp}^2) \delta (1-z_1) \delta (1-z_2). \nonumber
\end{eqnarray}
Here $\gamma_{J_{qq}}$, $\gamma_{J_{\bar{q}\bar{q}}}$  and $\gamma_S$ are the anomalous dimensions for 
$J_{qq}$, $J_{\bar{q}\bar{q}}$ and the soft function, which are given as
\begin{eqnarray}
\gamma_J (x,\mathbf{q}_{\perp}^2,\mu) &=& \frac{d}{d\ln \mu} J(x,\mathbf{q}_{\perp},\mu) = 
\frac{\alpha_s C_F}{2\pi} \frac{\delta (\mathbf{q}_{\perp}^2)}{\pi}
 \Bigl[2\delta (1-x) \ip -\frac{1+x^2}{(1-x)_+}\Bigr], \nonumber \\
\gamma_S (\mathbf{q}_{\perp}^2,\mu) &=& \frac{d}{d\ln \mu} S(\mathbf{q}_{\perp},\mu) = -2\frac{\alpha_s C_F}{\pi} 
\frac{\delta (\mathbf{q}_{\perp}^2)}{\pi} \ln \frac{\mu^2}{\Delta_1 \Delta_2},
\end{eqnarray}
where $\Delta_1 =p_1^2/\overline{n}\cdot p_1$, and $\Delta_2 = p_2^2/n\cdot p_2$,  and $J=J_{qq} =J_{\bar{q}\bar{q}}$. 
Note that $\gamma_J$ and $\gamma_S$
still depend on the IR-dependent terms, but the sum $\gamma_W$ has no  IR divergence.  Compared to 
the  anomalous dimensions corresponding to $\gamma_J$ and $\gamma_S$ in Ref.~\cite{Chiu:2012ir}, the dependence on the
rapidity scale $\nu$ is tantamount to the IR dependence in our approach.

The hard coefficient $C(Q,\mu)$ is given by
\begin{equation}
C(Q,\mu) =1+\frac{\alpha_s C_F}{4\pi}\Bigl (-\ln^2  \frac{\mu^2}{-Q^2} -3\ln \frac{\mu^2}{-Q^2} -8+\frac{\pi^2}{6}\Bigr),
\end{equation}
and the anomalous dimension of the hard part $H(Q,\mu) =|C(Q,\mu)|^2$ is given as
\begin{equation}
\gamma_H (\mu) = -\frac{\alpha_s C_F}{\pi} \Bigl( 2\ln \frac{\mu^2}{Q^2} +3\Bigr). 
\end{equation}
Combining $\gamma_W$, $\gamma_H$ and $\gamma_{\phi}$, it is clear that the factorized scattering cross section, 
Eq.~(\ref{finxsec}), is independent
of the renormalization scale $\mu$ at one loop.  The numerical analysis for the scattering cross sections is interesting with the solution
for the renormalization group equation.This can be applied to the processes with small transverse momentum such as Drell-Yan process,
Z boson production as well as Higgs production both at Tevatron and LHC. 
The numerical estimates involving TMDPDF in these processes  will be 
presented in future publication.

\section{Decoupling of the UV and IR divergences\label{decuvir}}
In order to see how the decoupling of the UV and IR divergences is achieved in various combinations, it is convenient to collect all 
the results obtained so far. The results for the integrated PDF are summarized as follows:
The collinear parts are separated into the virtual and real parts as
\begin{eqnarray}
M_{\mathrm{PDF}}^V (p_1^2)&=& \frac{\alpha_s C_F}{2\pi} \delta (1-x) \Bigl[\frac{2}{\epsilon^2} +\frac{3}{2\epsilon} +\frac{2}{\epsilon}
\ln \frac{\mu^2}{-p_1^2} +\frac{3}{2} \ln \frac{\mu^2}{-p_1^2} +\ln^2 \frac{\mu^2}{-p_1^2} +\frac{7}{2} -\frac{\pi^2}{6} \Bigr], 
\nonumber \\
M_{\mathrm{PDF}}^R (p_1^2) &=& \frac{\alpha_s C_F}{2\pi}\Bigl[\delta (1-x) \Bigl( -\frac{2}{\epsilon^2} -\frac{2}{\epsilon} 
\ln \frac{\mu^2}{-p_1^2} -\ln^2 \frac{\mu^2}{-p_1^2} -\frac{\pi^2}{6}\Bigr) -2 (1-x) \nonumber \\
&&+\frac{1+x^2}{(1-x)_+} \Bigl(\frac{1}{\epsilon} -\ln x +\ln \frac{\mu^2}{-p_1^2}\Bigr) +(1+x^2) \Bigl( \frac{\ln (1-x)}{1-x}\Bigr)_+\Bigr].
\end{eqnarray}
Note that there is another collinear contribution from the $\overline{n}$ direction, which is obtained from 
replacing $p_1^2$ by $p_2^2$. Similarly, the soft function is given as
\begin{eqnarray}
M_{\mathrm{soft}}^V = \frac{\alpha_s C_F}{2\pi} \Bigl( -\frac{2}{\epsilon^2} -\frac{2}{\epsilon} 
\ln \frac{\mu^2}{\Delta_1 \Delta_2}
-\ln^2 \frac{\mu^2}{\Delta_1 \Delta_2} -\frac{\pi^2}{2}\Bigr), \\
M_{\mathrm{soft}}^R = \frac{\alpha_s C_F}{2\pi} \Bigl( \frac{2}{\epsilon^2} +\frac{2}{\epsilon} 
\ln \frac{\mu^2}{\Delta_1 \Delta_2}
+\ln^2 \frac{\mu^2}{\Delta_1 \Delta_2} +\frac{\pi^2}{2}\Bigr). \nonumber
\end{eqnarray}
The mixing terms $(\ln \mu^2/(-p^2))/\epsilon$  exists in $M_{\mathrm{PDF}}^V$, 
$M_{\mathrm{PDF}}^R$, $M_{\mathrm{soft}}^V$ and $M_{\mathrm{soft}}^R$ respectively. 
However, if we add the collinear parts $M_{\mathrm{PDF}}^V+M_{\mathrm{PDF}}^R$, the mixing
term cancels. It is also true for the soft parts $M_{\mathrm{soft}}^V+M_{\mathrm{soft}}^R$. Therefore 
the total contribution is free of the mixing.
There is another combination in which the mixing cancels. They are 
$M_{\mathrm{PDF}}^V (p_1^2) +M_{\mathrm{PDF}}^V (p_2^2) +M_{\mathrm{soft}}^V$, and 
$M_{\mathrm{PDF}}^R (p_1^2) +M_{\mathrm{PDF}}^R (p_2^2) +M_{\mathrm{soft}}^R$. It means that the 
virtual corrections of the collinear and soft parts
do not contain the mixing of the UV and IR divergences, nor the real gluon emissions of the collinear and soft parts.

In the case of the TMDPDF, the virtual corrections of the collinear and soft parts have the same expression except $\delta (\kp)$, 
hence the mixing is not there. In the real gluon emissions, the possible mixing terms appear when $\kp$ approaches infinity 
(or $\Lambda_T^2$) in the terms
\begin{eqnarray} \label{nomix}
M_{\mathrm{TMDPDF}}^R &=& 2\tilde{M}_b +M_c \rightarrow 
-2\Bigl(\ln \frac{\mu^2}{-p_1^2} +\ln \frac{\mu^2}{-p_2^2}\Bigr) \Bigl[\frac{1}{\kp}\Bigr]_{\mu^2} \nonumber \\
M_{\mathrm{soft}}^R &=& M_a^s \rightarrow 2\Bigl(\ln \frac{\mu^2}{-\Delta_1} +\ln \frac{\mu^2}{-\Delta_2}\Bigr) 
\Bigl[\frac{1}{\bta^2}\Bigr]_{\mu^2},
\end{eqnarray}
where $\Delta_1= p_1^2/\overline{n}\cdot p_1$ and $\Delta_2 =p_2^2/n\cdot p_2$. Other terms may contain divergences, but the
UV and IR divergences are decoupled.
As can be seen clearly, these are cancelled
when the real gluon emissions in the collinear and soft parts are added. The mixing of the UV and IR divergences disappears
in the virtual corrections and in the real gluon emissions of the collinear and soft parts respectively. 

This observation is useful since the virtual corrections of the TMDPDF have the same form as those of the integrated PDF except 
$\delta (\kp)$. And they do not have the mixing term. Therefore the real gluon emissions should not have any mixing term either, 
when integrated over the transverse momentum. As can be seen in Eq.~(\ref{nomix}), the mixing term is cancelled before the 
integration over the transverse momentum. 
If it is true to all orders in $\alpha_s$, we can safely use this fact in order to obtain the anomalous dimension of the combined
collinear and soft parts.

\section{Conclusions\label{conc}}

The divergence structure in the TMDPDF is intricate because the rapidity divergence should be also handled as well as the
usual UV and IR divergences. One of the major problems in studying TMDPDF is to find a regularization method to treat 
the rapidity divergence. The dimensional regularization can handle the UV divergence, and possibly the IR divergence, but it cannot
isolate the rapidity divergence. As far as the UV and IR divergences are concerned, it is convenient to use dimensional regularization
to extract the UV divergence, and to employ the offshellness of external particles. There are various ways to treat rapidity divergence, 
and we have found that a simple way to regulate the rapidity divergence is achieved by inserting the $\delta$ regulators in 
the collinear Wilson lines only.  Individual collinear Feynman diagrams 
depend on the regulator $\delta$, but the total collinear contribution with the zero-bin subtraction 
is independent of the regulator, hence no rapidity divergence. The regulators similar to $\delta$ may be inserted, 
say, in the soft Wilson lines, or in the collinear fermion
propagator, but they regulate the IR divergence, not the rapidity divergence. 
  
For comparison, let us recapitulate previous approaches in handling the rapidity divergence. 
Collins and Soper \cite{Collins:1981uk} tilted the lightcone such that the rapidity divergence is handled by an extra scale in
association with this tilting. Chiu et al. \cite{Chiu:2012ir} modified the collinear and soft Wilson lines by introducing a rapidity 
scale $\nu$, which treats the rapidity divergence. In both approaches, the hard, collinear and soft parts depend on the additional scale,
while the cross sections do not. Therefore another renormalization group equation with respect to this scale is introduced. Another
approach is to choose the axial gauge \cite{Belitsky:2002sm,Cherednikov:2007tw}, for example $\overline{n}\cdot A_n=0$, 
instead of covariant gauges such as the Feynman gauge 
we chose. The advantage of this gauge choice is that many Feynman diagrams from the Wilson lines vanish, but there should be 
additional transverse gauge link which links the operators for TMDPDF at infinity so that it becomes gauge invariant.  
Ref.~\cite{GarciaEchevarria:2011rb} treats the rapidity divergence in a similar way to ours, but they insert  the $\delta$ regulators
in both the collinear and the soft Wilson lines and give a specific relation between the collinear and the soft regulators. That relation,
as we point out, holds only for the back-to-back current.

The separation of the UV and IR divergences is achieved in the sum of the virtual and real corrections in the collinear sector
and in the soft sector, and in the sum of the collinear and soft parts in the virtual corrections and in the real gluon emissions.
In effective theories like SCET, the mixing shows up often in intermediate steps, but it cancels in the final result. In the 
radiative correction of the TMDPDF, it also holds in the sums mentioned above.  Therefore the renormalization
group scaling can be considered only for the hard part and the sum of the collinear and soft parts. This prescription seems to violate
the factorization in its strictest sense that the collinear and the soft parts are separately well defined. 
We rather claim that the  attempt to combine the collinear part with part of the soft part such that the UV and IR divergences 
are decoupled is futile in its strictest sense that no such thing occurs with general
regulators unless a very specific set of relations is imposed. The soft part interacts with all the collinear sectors in the process, hence
contains the information of all the collinear sectors through their offshellness.  Therefore it is physically more reasonable 
to consider the scaling behavior of the sum of the collinear and soft parts in order to decouple the UV and IR divergences. 
This may sound cumbersome, since it means that
we have to compute all the collinear and soft parts first  in complicated processes with multijets or heavy 
colored particles.  But we stress that it is the correct procedure to treat
high-energy scattering. It will be interesting to see if our procedure also works in 
other processes involving small transverse momentum. Various aspects of our formalism will be 
investigated in future publications.

The  important result in this work is the factorization formula for the differential scattering cross section in Drell-Yan process
with small transverse momentum, given by Eq.~(\ref{vir}). The scattering cross section is factorized into the hard part, the TMD
kernel and the PDF's. The hard part is the matching coefficient between the full theory  and the SCET with the scale $Q\lambda$. 
The TMD kernel is obtained by matching the two SCET's with the scales $Q\lambda$ and $Q\lambda^2$. Therefore the TMD
kernel should be free from IR divergence to all orders, and it is shown explicitly at NLO. Note that the TMD kernel 
$W(z_1, z_2, \mathbf{q}_{\perp},\mu)$ is expressed in momentum space, rather than in impact parameter space. At least at 
NLO, the anomalous dimension of $W$ is proportional to $\delta (\mathbf{q}_{\perp}^2)$, hence the dependence of the 
scaling behavior on the transverse momentum is trivial.

\begin{acknowledgments}
J.~Chay was supported by the National Research Foundation of Korea (NRF) grant funded by the Korea government (MEST) No.
2009-0086383. C.~Kim was supported by Basic Science Research Program through the National Research Foundation of Korea (NRF)
funded by the Ministry of Education, Science and Technology (No. 2012R1A1A1003015). 
\end{acknowledgments}

\appendix*
\section{$\mu^2$ distribution functions and infinity distribution functions}


Suppose that a function $g(x)$ diverges as $x\rightarrow 0$. Then the plus 
distribution function $[g(x)]_b$ can be defined, in general, as
\begin{eqnarray} \label{genplus}
\int_0^a [g(x)]_b f(x) dx &=&\int_0^a dx g(x) f(x) -f(0) \int_0^b dx g(x) \nonumber \\
&=& \int_0^a g(x) \Bigl[f(x) -f(0)\Bigr] - f(0) \int_a^b dx g(x),
\end{eqnarray}
for a regular test function $f(x)$. The need for the plus distribution function is to separate the IR divergent part. 
When $g(x)$ diverges as $x\rightarrow 0$,  the integral is tamed to be finite  by subtracting $f(0)$.  In this general 
definition of the plus distribution function, the upper limits $a$ and $b$ are arbitrary as long as the lower limits
of the two integrals in Eq.~(\ref{genplus})  are both zero to guarantee that the IR divergence at $x=0$ is cancelled in this subtraction.
The finite remainder, the second term in Eq.~(\ref{genplus}), depends on the limits $a$ and $b$.  
In a special case $a=b=1$, the familiar plus distribution function is obtained as
\begin{equation}
\int_0^1 [g(x)]_+ f(x) =\int_0^1 dxg(x) \Bigl[f(x) -f(0)\Bigr].
\end{equation}

There are two types of functions in the radiative corrections, which are of the form
\begin{equation}
g_1 (\kp, \epsilon, \delta) = \frac{\mu^{2\epsilon}}{(\kp+\Delta)^{1+\epsilon}}, \ g_2 (\kp, \epsilon, \delta)
= \frac{\mu^{2\epsilon}  \displaystyle \ln \frac{\kp}{\Delta}}{(\kp -\Delta)^{1+\epsilon}},
\end{equation}
where $\Delta$ approaches zero. If $\Delta=0$, these functions are singular at $\kp=0$, that is, IR divergent. We extract the IR
divergence as the coefficient of  $\delta (\kp)$, and the remainder is the distribution function free of the divergence. 
With the presence of the nonzero $\Delta$,  the IR divergence is regulated by $\Delta$ instead. However, there is a big difference in
considering the distribution functions $g_1$ and $g_2$ in $\kp$-space compared to $[g(x)]_b$ since the upper 
limit can reach infinity. Irrespective of whether the upper limit goes to infinity or not, $g_1$ and $g_2$ diverge at $\kp=0$ with
$\Delta =0$. If the upper limit in considering the integral including $g_1$ or $g_2$ reaches infinity, the integral also has an UV
divergence. Therefore we classify the distribution functions according to the upper limits. These two cases have their own physical
importance. If we restrict the transverse momentum to be of order $Q\lambda$ and collect experimental data, there is finite upper
limit which we call $\Lambda_T$.  The relevant distribution function is called the ``$\mu^2$ distribution". In the effective theory, 
$\kp$ can reach infinity and the UV divergence from the integration offers the information on the scaling behavior. The ``infinity
distribution" is devised for this case.

First let us consider the case in which the transverse momentum remains finite up to 
some scale $\Lambda_T$. 
If a regular test function $f(\kp)$ multiplied by $g_1(\kp, \epsilon, \Delta)$ is integrated from 0 to $\Lambda_T$, 
it yields
\begin{eqnarray} \label{testf}
&& \int_0^{\Lambda_T^2} d\kp g_1 (\kp,\epsilon,\Delta) f(\kp)  \nonumber \\
&&=\int_0^{\Lambda_T^2} d\kp g_1(\kp,\epsilon,\Delta) f(\kp) -\int_0^{\mu^2} d\kp g_1(\kp,\epsilon,\Delta) f(0)  
+\int_0^{\mu^2} d\kp g_1(\kp,\epsilon,\Delta) f(0)  \nonumber \\
&&= \int_0^{\Lambda_T^2} d\kp g_1 (\kp,0,0) f(\kp) -\int_0^{\mu^2} d\kp g_1(\kp,0,0) f(0) 
+f(0)\ln \frac{\mu^2}{\Delta}\nonumber \\
&&\equiv \int_0^{\Lambda_T^2} d\kp  \Bigl[g_1(\kp,0,0)\Bigr]_{\mu^2} f(\kp) +f(0)\ln \frac{\mu^2}{\Delta}.
\end{eqnarray}
In the final expression,  we put $\epsilon=0$, and $\Delta =0$ in $g(\kp, \epsilon,\Delta)$  because 
the integral  has no pole in $\epsilon$, and
the possible infrared divergence near $\kp \sim 0$ is cancelled in the subtraction.  This defines the $\mu^2$ distribution. $g_2$ can
be treated in a similar way to yield
\begin{eqnarray}
g_1(\kp, \epsilon, \Delta) &=& \frac{\mu^{2\epsilon}}{(\kp+\Delta)^{1+\epsilon}} =\delta (\kp) \ln \frac{\mu^2}{\Delta}
+\Bigl[\frac{1}{\kp}\Bigr]_{\mu^2}, \nonumber \\
g_2(\kp, \epsilon, \Delta) &=& = \frac{\mu^{2\epsilon}  \displaystyle \ln \frac{\kp}{\Delta}}{(\kp -\Delta)^{1+\epsilon}}
=\delta (\kp) \Bigl( \frac{1}{2} \ln^2 \frac{\mu^2}{\Delta} +\frac{\pi^2}{3}\Bigr) \nonumber \\
&&+\Bigl[\frac{\ln \kp/\mu^2}{\kp}\Bigr]_{\mu^2}
+\Bigl[\frac{1}{\kp}\Bigr]_{\mu^2} \ln\frac{\mu^2}{\Delta}.
\end{eqnarray}
Note that these functions are independent of $\mu$ since we can put $\epsilon=0$ from the beginning.  However, 
the $\mu$ dependence is split into the delta function and the $\mu^2$ distribution function for convenience.
The independence can be seen by computing the integral
\begin{eqnarray}
&& \int_0^{\Lambda_T^2} d\kp g_1(\kp,\epsilon,\Delta) f(\kp)  \nonumber \\
&&= \int_0^{\Lambda_T^2} d\kp  \Bigl[g_1(\kp,0,0)\Bigr]_{\mu^2} f(\kp) +f(0)\ln \frac{\mu^2}{\Delta}  \nonumber \\
&&= \int_0^{\Lambda_T^2} d\kp g_1(\kp,0,0) \Bigl[ f(\kp) -f(0)\Bigr] +f(0) \ln \frac{\Lambda_T^2}{\Delta}. 
\end{eqnarray}

If the upper limit for the transverse momentum reaches infinity, $g_1$ and $g_2$ contain UV
divergences:
\begin{eqnarray}
\int_0^{\infty} d\kp g_1(\kp,\epsilon,\Delta) &=& \frac{1}{\epsilon} +\ln \frac{\mu^2}{\Delta} +\mathcal{O}(\epsilon)  
\nonumber \\
\int_0^{\infty} d\kp g_2(\kp,\epsilon,\Delta) &=&\frac{1}{\epsilon^2} +\frac{1}{\epsilon}\ln \frac{\mu^2}{\Delta} 
+\frac{1}{2}\ln^2 \frac{\mu^2}{\Delta} +\frac{\pi^2}{3} +\mathcal{O}(\epsilon) .
\end{eqnarray}
Therefore $g_1$ and $g_2$ can be expressed as
\begin{eqnarray} \label{infdis}
g_1 (\kp,\epsilon,\Delta) &=& \Bigl(\frac{1}{\epsilon} +\ln \frac{\mu^2}{\Delta}\Bigr) \delta (\kp) 
+\Bigl[g_1 (\kp,\epsilon,0)\Bigr]_{\infty}  \\
g_2 (\kp, \epsilon, \Delta) &=& 
\Bigl( \frac{1}{\epsilon^2} +\frac{1}{\epsilon}\ln \frac{\mu^2}{\Delta} +\frac{1}{2}\ln^2 \frac{\mu^2}{\Delta} +\frac{\pi^2}{3}\Bigr)
\delta (\kp) +\Bigl[\frac{1}{\kp}\Bigr]_{\infty} \ln \frac{\mu^2}{\Delta} +\Bigl[\frac{\ln \kp/\mu^2}{\kp}\Bigr]_{\infty}.
\nonumber 
\end{eqnarray}
Note that Eq.~(\ref{infdis}) is not strictly correct since the UV divergence is not peaked near $\kp \sim 0$. Therefore they are meaningful
only inside the integral over the transverse momentum which reaches infinity. Then the UV divergence is correctly accounted for.

\end{document}